\documentclass[pdflatex,sn-mathphys-num]{sn-jnl}

\usepackage{graphicx}%
\usepackage{multirow}%
\usepackage{amsmath,amssymb,amsfonts}%
\usepackage{amsthm}%
\usepackage{mathrsfs}%
\usepackage[title]{appendix}%
\usepackage{xcolor}%
\usepackage{textcomp}%
\usepackage{manyfoot}%
\usepackage{booktabs}%
\usepackage{algorithm}%
\usepackage{algorithmicx}%
\usepackage{algpseudocode}%
\usepackage{listings}%

\def\b{\begin{equation}}
    \def\e{\end{equation}}

   \newcommand{\dif}{\mathrm{d}}
   \newcommand{\bq}{\begin{eqnarray*}}
   \newcommand{\eq}{\end{eqnarray*}}
   \newcommand{\beq}{\begin{eqnarray}}
   \newcommand{\enq}{\end{eqnarray}}

\theoremstyle{thmstyleone}%
\theoremstyle{thmstyletwo}%

\theoremstyle{thmstylethree}%

\begin{document}

\title{Non-perturbative approach for scalar particle production in Higgs-$R^2$ inflation}


\author*[1]{\fnm{Flavio} \sur{Pineda}}\email{fpineda@xanum.uam.mx}

\author[1]{\fnm{Luis O.} \sur{Pimentel}}\email{lopr@xanum.uam.mx}
\equalcont{These authors contributed equally to this work.}

\affil*[1]{\orgdiv{Departamento de Física}, \orgname{Universidad Autónoma Metropolitana}, \orgaddress{\street{Av. Ferrocarril San Rafael Atlixco 186, P. O. Box 55-534}, \city{Iztapalapa}, \postcode{09310}, \state{CDMX}, \country{México}}}


\abstract{Gravitational particle production in cosmology is a mechanism through which particles of different natures are produced during the very early universe. It is a general mechanism that explains how the universe became populated with the particles of the Standard Model after cosmic inflation and may also account for the origin of dark matter. In this work, we study the non-perturbative production of massive scalar particles in the Higgs-$R^2$ inflation model, a two-field scalar inflation model within the Einstein frame. We consider spectator scalar fields that are conformally or minimally coupled to the gravitational field through the curvature scalar $R$ which in turn is a time-dependent function determined by the fields driving inflationary dynamics. We numerically compute the production of these particles using the Bogolyubov transformation method for each scenario, aiming to assess the spectrum of the produced particles. For both scenarios, we consider light particles with masses $m_\chi \ll H_\mathrm{end}$ and large masses  that exceed the Hubble scale at the end of inflation $m_\chi \gtrsim H_\mathrm{end}$. We use these numerical results to calculate the relic abundance $\Omega_\chi h^2$ to find out if the model is viable as a dark matter candidate.}


\maketitle

\section{Introduction}\label{sec1}

Cosmic inflation is the cornerstone of modern cosmology, providing a framework to understand the early universe and predict its fundamental characteristics. 
It describes a period of accelerated expansion driven by a scalar field \(\phi\) that slowly rolls towards the minimum of its potential \(V(\phi)\). This mechanism addresses classical problems of the standard cosmological model and has become a central paradigm in cosmology \cite{Guth1981, LINDE1982389}. During the inflationary epoch, significant particle production occurs, particularly when the kinetic energy of the inflaton field becomes comparable to its potential energy. This particle production can arise from gravitational effects due to the universe's expansion and the dynamics of the inflaton field, serving as a classical background for propagating quantum fields \cite{Parker1969, Ford1987}. This phenomenon, known as gravitational particle production, is a fundamental mechanism in quantum field theory in curved spacetime, offering insights into the origins of dark matter and the preheating and reheating epochs following inflation \cite{kolb2023}. Several scenarios have been proposed for vacuum particle production during the early universe, particularly during inflation due to the evolution of the inflaton field and the universe's expansion. Typically, the inflaton field is coupled to matter fields (bosonic, fermionic, or gauge fields) by a coupling constant \(g\). At a critical value \(\phi = \phi_*\) of the inflaton field, resonance is triggered, leading to instantaneous particle production \cite{Chung2000, romano2008}. However, matter fields need not be directly coupled with the inflaton or standard model particles. Massive particles can be produced purely through gravitational effects, either via direct couplings with the gravitational field or induced couplings through a conformal transformation. If the contribution of the matter field to the energy density is negligible, the inflationary dynamics remain unaffected, leading to no back-reaction effects on spacetime geometry. Such fields are called spectator fields, and this mechanism provides a natural explanation for the origin of dark matter \cite{Gorbunov2012, ema2018production, Chung2003, Chung2001, Hashiba2019, cembranos2020}. Beyond its role in the origin of dark matter, gravitational particle production is also linked to key early-universe phenomena such as preheating after inflation \cite{kofman1997, kaiser1997, Shtanov1995, dorsch2024}, primordial gravitational waves \cite{Adshead2019, Cook2012, Garcia-Bellido2011}, and primordial magnetogenesis \cite{Durrer2013, Giovannini2000, Kobayashi2014, kandus2011primordial}. Furthermore, it leaves significant observable imprints on the Cosmic Microwave Background (CMB), particularly in the generation of primordial non-Gaussianities \cite{Barnaby2012, Barnabywaves2012, langlois2009primordial}.

In this work, we study gravitational particle production during and immediately after inflation driven by the SM Higgs boson and the scalaron field in the Higgs-\(R^2\) inflation model. This inflation model extends the SM Higgs inflationary model \cite{Bezrukov2009} to address the problem of large coupling \cite{GORBUNOV2019, EMA2017}. Here, cosmic inflation is driven by the Higgs field and the extra scalar degree of freedom from the quadratic \(R\) term, known as the scalaron. Unlike other multi-field inflation models, the Higgs-\(R^2\) model effectively reduces to a single-field model during inflation, with the Higgs field acting as an auxiliary field while the scalaron drives inflation \cite{GUNDHI2020114989}. At the end of inflation during the reheating phase, both fields and the Ricci scalar oscillate around the potential minimum. This oscillatory behavior enhances scalar particle production through non-perturbative mechanisms such as tachyonic instabilities \cite{bassett1998, dufaux2006preheating}. The preheating phase after inflation has been studied \cite{He2019, He2021, He2021tachionic, BEZRUKOV2019657}, as well as the production of primordial magnetic fields during the inflationary phase \cite{Durrer2022}, considering effective single-field inflation. We propose an extension to the Higgs-\(R^2\) model by introducing a massive scalar field \(\chi\) as a spectator field that does not contribute to the energy density during the inflationary epoch, with negligible interactions with the SM but considering a direct coupling to the Ricci scalar via a non-minimal coupling term \(\xi_\chi R\,\chi^2\). Initially, the inflationary sector and the spectator field are not explicitly coupled in the Jordan frame; however, gravitational interactions between the scalaron, the Higgs field, and free scalars induce couplings upon transforming to the Einstein frame. This process results in gravitational particle production with significant implications for preheating after inflation and background dynamics. Induced coupling in the Einstein frame between free scalars and the scalar sector of the action provides an excellent channel for scalaron decay into light scalars. However, in the non-perturbative approach, the field mass \(m_\chi\) does not necessarily have to be light, and we can explore the possibility of scalar particle production for spectator field masses $m_\chi \gtrsim H_\mathrm{end}$, where $H_\mathrm{end}$ is the Hubble ratio at the end of inflation. We take into account the oscillatory behavior of background dynamics after inflation as a classical background, which results in particle production due to the time-dependent background inducing a time-dependent frequency in the mode equations for the spectator field \(\chi\). To achieve this, we employ the Bogoliubov approach \cite{Parker1969} to compute the comoving number density \(n_k(t)\) of particles generated by the background dynamics in Higgs-\(R^2\) inflation, solving the mode equations for the quantized modes \(\chi\).

The article is organized as follows: Section 2 provides a brief introduction to the Higgs-\(R^2\) model and the background dynamics during and after inflation for a set of suitable initial conditions. Section 3 analyzes the couplings induced by gravitational effects in the Einstein frame. Section 4, we derive the mode equation for the spectator field, which is necessary to investigate the behavior of the time-dependent frequency during and after inflation. We adopt a numerical approach to solve the mode equation for modes that leave 5 e-folds before the horizon crossing during inflation, ensuring that each mode is within the horizon at the start of the numerical simulation. Section 4 examines the production of massive particles during preheating. In section 5 we show the numerical results for the gravitational production spectra for $\xi_\chi =0$ and $\xi_\chi = 1/6$ and for the time-evolution of $\beta_k$-Bogolyubov coefficient. Lastly, section 6 is dedicated to the relic abundance of gravitational production of scalars.

In the following, we will use natural units in which $\hbar = 1$, $c = 1$, $\kappa_\mathrm{B} = 1$ and we use the mass Planck as $8\pi G = M_p^{-2}$. We also work in the context of a spatially flat, homogeneous, and isotropic universe described by the FLRW metric

\beq 
\nonumber \dif s^2 &=& -\dif t^2 + a^2(t)(\dif x^2 + \dif y^2 + \dif z^2)\\\\
\label{FLRW metric}
\nonumber &=& a^{2}(\tau)(-\dif \tau^2 + \dif x^2 + \dif y^2 + \dif z^2)\,,
\enq 

where $t$ is the cosmic time, $\tau$ the conformal time and the relation between the two is $\dif \tau = \dif t/a(t)$ with $a(t)$ is the scale factor.


\section{Brief review of Higgs-$R^2$ inflationary model}\label{sec2}

In this section, we will give a brief review of the Higgs-$R^2$ model in both Einstein and Jordan frames. For further details, we recommend the main references \cite{GUNDHI2020114989, He_2018, EMA2017, GORBUNOV2019} and the references therein. Let us get started with the discussion in the Jordan frame; the model is given by the action

\beq 
    S_\mathrm{J} = \int \dif x^4 \sqrt{-g_J}\left[\dfrac{1}{2}(M_p^2 + \xi_h h^2)R_J +\dfrac{\xi_s}{4}\,R_J^2 -\dfrac{1}{2}g^{\mu\nu}_J\,\partial_\mu h\,\partial_\nu h -\dfrac{\lambda}{4}\,h^4 \right]\,,
    \label{action jordan}
\enq 
 
 where $h$ is the Higgs field in the unitary gauge, the subindex $J$ denotes variables defined in the Jordan frame, $\xi_h$ is the constant coupling between the Higgs field and the gravitational sector, $\xi_s$ is a mass parameter and $\lambda $ is the quartic self constant coupling of the Higgs field. This model contains a quadratic term in $R_J$ described by the constant $\xi_s$, which comes from the $R^2$ model \cite{STAROBINSKY198099}. Such a constant is related to the scalaron mass as $\xi_s = M_p^2/3M^2$.  The model \eqref{action jordan} can be regarded as a UV extension of Higgs inflation, so it is interesting to make a double interpretation of the gravitational coupling in this conformal frame. In Higgs inflation, the constant coupling $\xi_h$ must fulfill the condition $\xi_h^2/\lambda \sim 10^{10}$ to be consistent with CMB observations \cite{Bezrukov2009, barvinsky2008}. This causes the cutoff scale of the model to be below the natural energy scale (Planck scale) during inflation, making the predictions of the model questionable; however, it has been shown that including the Ricci scalar quadratic term modifies the value of the energy scale, raising it to the Planck scale. To solve the strong coupling problem present in the scalar sector, the parameter $\xi_s$ has to obey the relation \cite{GORBUNOV2019}

\beq 
    \xi_s \gtrsim \dfrac{\xi_h^2}{4\pi}\,.
\enq 

This expression allows us to make a perturbative treatment of the model. To describe the inflationary dynamics of the model, it is more convenient to bring the action \eqref{action jordan} into the canonical Einstein-Hilbert form. So we need to carry out a conformal transformation $\Psi^2(x)$ introducing an extra scalar degree of freedom $\phi$ called scalaron, which comes from the quadratic term in $R_J$. Thus, to go from the Jordan frame to Einstein frame one has to perform a conformal transformation of the space-time metric $g_{\mu\nu}\to \Psi^2(x)g_{\mu\nu}$ with $\Psi(x) > 0$, where the correct form of it is

\beq 
    \phi = \alpha^{-1}\log \Psi^2\,,
    \label{conformal transformation}
\enq 

where $\alpha = \sqrt{2/3}M_p^{-1}$. The conformal transformation \eqref{conformal transformation} removes the quadratic term of the scalar curvature $R_J$ introducing the scalaron $\phi$.  Then in the Einstein frame, the Higgs-$R^2$ model of inflation takes the form of a two-scalar fields model with a non-canonical kinetic term

\begin{figure}[h]
\centering
\includegraphics[width=0.48\textwidth]{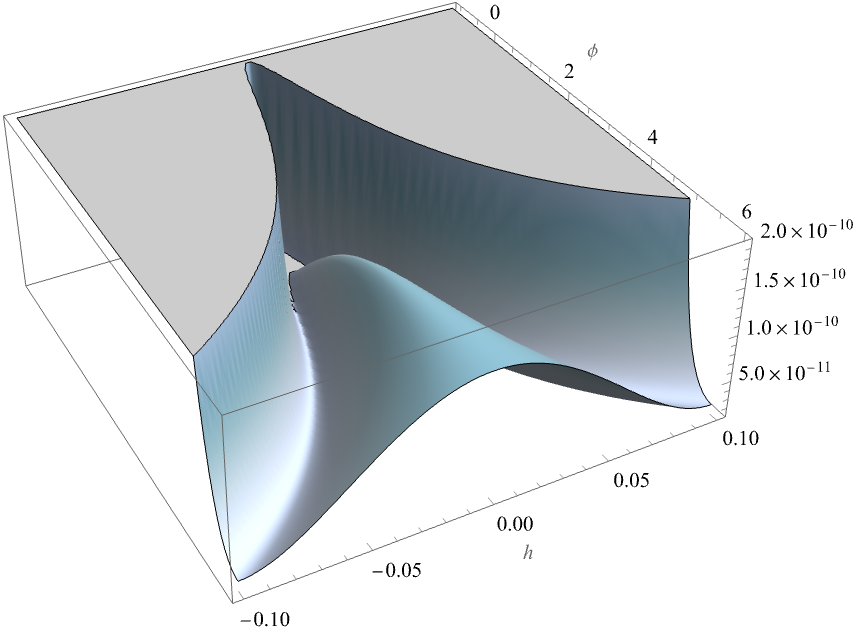} \quad \includegraphics[width=0.48\textwidth]{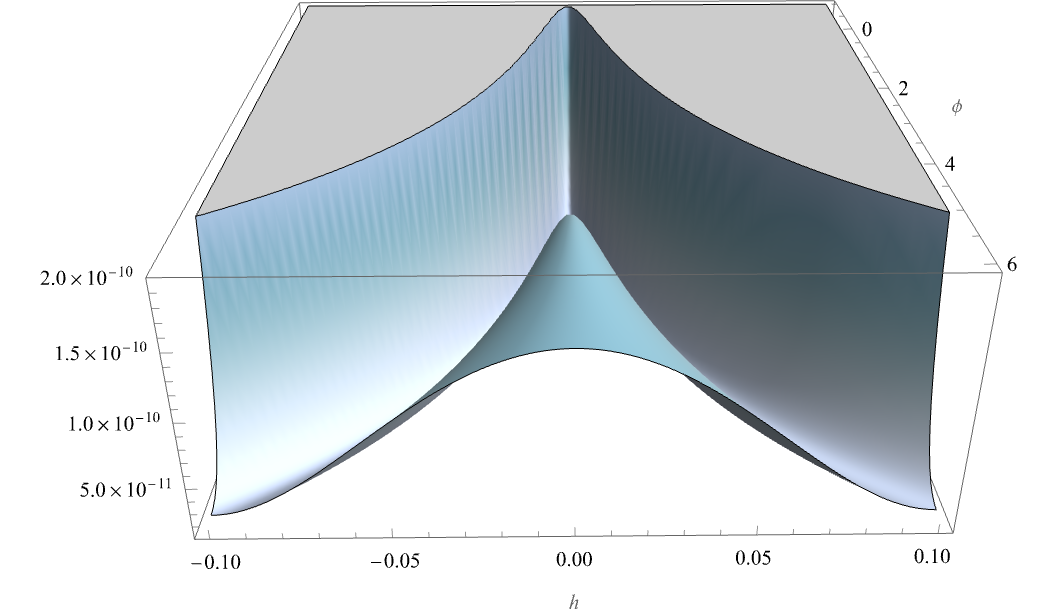}
\caption{Scalar potential in the Einstein frame of the Higgs-$R^2$ model seen from different angles. The curve $\phi = \mathbf{cte}$ has the shape of the potential of the $R^2$ inflation model, while $h = \mathrm{cte}$ has the shape of the quartic Higgs potential of the Standard Model. We have restricted ourselves to the standard parameter space $\xi_h = 4000$, $\xi_s = 2\times 10^9$ and $\lambda = 0.013$.}
    \label{fig1}
\end{figure}

\beq 
S_E = \int\dif^4 x\,\sqrt{-g}\,\left\{\dfrac{M_p^2}{2}R-\dfrac{1}{2}g^{\mu\nu}\partial_\mu\phi\,\partial_\nu\phi - \dfrac{1}{2}\,e^{-\alpha\phi}g^{\mu\nu}\partial_h\,\partial_\nu h-V(\phi\,,h)\right\}\,,
        \label{Action Einstein}
\enq

which can be written in the general action for multi-field inflationary models

\beq 
    S=\int \dif^4 x\,\sqrt{-g}\left[\dfrac{M_p^2}{2}R  -\dfrac{1}{2}G_{IJ}\,g^{\mu\nu}\partial_\mu \phi^I\,\partial_\nu\phi^J - V(\phi^I)\right]\,,
\enq 

where in our case $\phi^I = (\phi\,,h)$ and $G_{IJ}$ is the non-flat field-space metric given by 

\beq 
G_{IJ} = \begin{pmatrix}
    1 & 0\\
    0 & e^{-\alpha\phi}
\end{pmatrix}\,.
\enq

We can see that the non-canonical kinetic term is associated with the geometry of the field space via the metric $G_{IJ}$ \cite{KARAMITSOS2018219}. On the other hand, the potential in the Einstein frame has interaction terms between the Higgs field $h$ and the scalaron $\phi$ and takes the form

\beq 
\nonumber V(\phi\,,h) = \dfrac{1}{4}\,e^{-2\alpha\phi}\left[\dfrac{M_p^4}{\xi_s}\left(e^{\alpha\phi}-1-\dfrac{\xi_h}{M_p^2}\,h^2 \right)^2+\lambda\,h^4\right]\,.\\
\label{potential Einstein frame}
\enq 

The potential $V(\phi^I)$ depends on both the $\phi$ and $h$ and mixes terms which make it more difficult to analyze analytically the background trajectory.

Nevertheless, during inflation, we can make an effective approximation of the model such that inflation is driven by the single scalar field $\phi$ \cite{GUNDHI2020114989} and with the following interesting features 

\begin{enumerate}
    \item $V(\phi^I)$ shows two valleys and one ridge.
    \item For  $\xi_h$, $\xi_s$ positive and enough large, the potential presents an attractor behavior.
    
    \item If $\phi \gg M_p$ with arbitrary initial conditions, the fields $\phi$ and $h$ will always evolve towards one of the valleys thus beginning the slow-roll evolution.
    
    \item If $\phi \ll M_p$, the potential is reduced to Higgs potential 
    
    \beq 
        V(\phi \ll M_p\,,h)\simeq V(h) = \dfrac{\lambda_\mathrm{eff}}{4}\,h^4\,,
    \enq 

    where $\lambda_\mathrm{eff}$ is the effective quartic self-coupling constant defined by
    
    \beq 
    \lambda_\mathrm{eff} = \lambda + \dfrac{\xi_h^2}{\xi_s}\,.
    \enq 

    The condition $\lambda_\mathrm{eff} \lesssim 4\pi$ is necessary for perturbative preheating

    \item If $h\ll M_p$, the potential is reduced to Starobinsky potential
    
    \beq
        V(\phi\,,h \ll M_p) \simeq \dfrac{M_p^4}{4\xi_s}(1-e^{-\alpha\phi})^2\,.
   \enq  
   
\end{enumerate}

We can see the positions of the valleys and the hill at $h=0$ in Fig. \ref{fig1}. The inflationary dynamics occur along one of these valleys, showing a universal attractor behavior, so that inevitably the fields $\phi$ and $h$ will always fall into one of the valleys and then drive slow-roll inflation. To find the valleys, we can perform the analysis in both Einstein \cite{GUNDHI2020114989} and Jordan \cite{PhysRevD.89.043527}  frames. In the Einstein frame in the limit $\phi\gg M_p$ (it is required for enough long inflation) and $h \gg v$, where $v = 246\,\mathrm{GeV}$, the valleys are found out by the condition 

\beq 
    \dfrac{\partial V}{\partial h} = 0\,,
    \label{valleys condition}
    \enq 

which implies that the Higgs field is completely determined by the inflaton field $\phi$

\beq 
    h^2_\mathrm{min}(\phi) = \dfrac{M_p^2\xi_h}{\xi_h^2+\lambda\xi_s}\,(e^{\alpha\phi}-1)\,.
    \label{higgs field condition Einstein}
\enq 

Because the potential \eqref{potential Einstein frame} is symmetric with respect to transformation $h\to-h$, the two valleys are localized by the condition \eqref{higgs field condition Einstein}. Thus, this shows that the Higgs field is an auxiliary field during inflation, so if we put the result \eqref{higgs field condition Einstein} back into the Einstein potential \eqref{potential Einstein frame}, we give the effective potential along the valleys 

\beq 
    V_\mathrm{eff}(\phi\,,h_\mathrm{min}) = \dfrac{M_p^4}{4
    (\xi_s+\xi_h^2/\lambda)}\,(1-e^{-\alpha\phi})^2
    \label{effective potential}
\enq

Notice that the effective potential reduces to Higgs inflation if $\lambda\xi_s \ll \xi_h^2$ while one gets $R^2$ inflation for $\lambda\xi_s \gg \xi_h^2$. To provide the correct amplitude of primordial scalar perturbations, the amplitude of \eqref{effective potential} has to fulfill \cite{planck2018x}

\beq 
     \xi_s + \dfrac{\xi_h^2}{\lambda} = \dfrac{N_*^2}{72\pi^2A_\zeta}\simeq 2\times 10^9\,,
    \label{condition CMB}
\enq

where $N_* \sim 50-60$ is the number of e-folds before the end of inflation, and $A_\zeta \sim 2\times10^{-9}$ is the amplitude of the primordial power spectrum. For $\lambda = 0.013$ and $\xi_h\,,\xi_s>0$ which satisfy \eqref{condition CMB}, the inflationary dynamics is effectively the same as single-field slow-roll inflation. In this paper, we will only analyze the standard parameter region $\xi_h\,,\xi_s\,,\lambda >0$. Specifically, if we have $\xi_h \gg 0$ the isocurvature effects of the model are absent the bending is within the weak mixing case $\eta_\perp \to 0$ \cite{GUNDHI2020114989}, and the isocurvature and adiabatic modes are not coupled so that they evolve independently in the standard parameter region of Higgs-$R^2$ model. Therefore, the description of the dynamics inside the valleys is an effective single-field inflation.  

\subsection{Background dynamics}

What follows is to analyze the dynamics of the model. We assume that during inflation both scalaron $\phi$ and Higgs field $h$ are homogeneous and only depend on cosmic time $t$. In this case, the equations of motion can be written as

\beq 
    \ddot{\phi} + 3H\dot{\phi} + \dfrac{\partial V}{\partial \phi} &=& -\dfrac{\alpha}{2}\,e^{-\alpha \phi}\dot{h}^2\label{scalaron equation}\,,\\[0.3cm]
     \ddot{h} + 3H\dot{h} + e^{\alpha\phi}\,\dfrac{\partial V}{\partial h} &=& \alpha\dot{\phi}\,\dot{h}\,.
     \label{higgs equation}
\enq 

The Einstein equations give the Friedman equations for $H(t) = \dot{a}/a$

\beq
    3M_p^2\,H^2 &=& \dfrac{1}{2}\dot{\phi}^2 + \dfrac{1}{2}\,e^{-\alpha\phi}\dot{h}^2 + V(\phi\,,h) \label{1 Friedmann}\,,\\[0.2cm]
    \dot{H} &=& -\dfrac{1}{2M_p^2}\,\left(\dot{\phi}^2 + e^{-\alpha\phi}\dot{h}^2\right)\,.
    \label{2 Friedmann}
\enq 

\begin{figure}[h]
\centering
\includegraphics[width=0.8\textwidth]{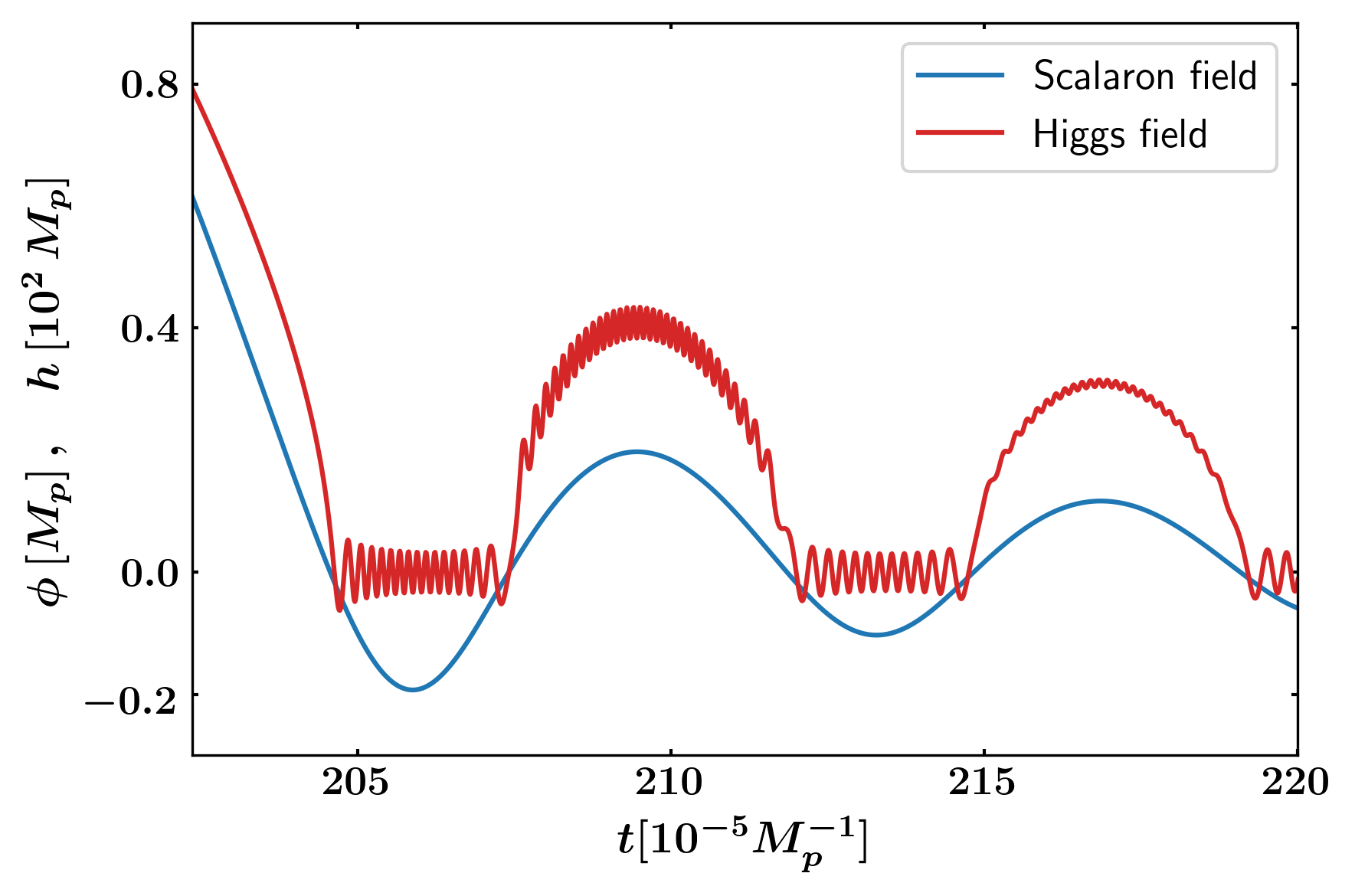}
\caption{We present the numerical solution of the background dynamics after inflation in cosmic time $t$ for the Higgs and scalaron fields within the valley of the potential. We have numerically solved \eqref{scalaron equation}, \eqref{higgs equation}, and \eqref{1 Friedmann} using the numerical parameters $\xi_h = 4000$, $\xi_s = 2\times 10^9$ and $\lambda = 0.013$ together with initial conditions $\phi(t_0) = 5.7 M_p$ and $h(t_0) = 0.096 M_p$.}
    \label{fig2}
\end{figure}

One has to numerically solve the system of equations \eqref{scalaron equation}, \eqref{higgs equation}, \eqref{1 Friedmann} and \eqref{2 Friedmann} to explore the inflationary dynamics of the model in the Einstein frame given some combination values of the parameters $\xi_h\,,\xi_s\,, \lambda$ and given suitable initial conditions for $\phi(t_0)$, $h(t_0)$. The inflation occurs when $\phi \gg M_p$ and the value of the auxiliary field $h$ is given by \eqref{higgs field condition Einstein} inside one of the valleys of the potential. In the Fig. \ref{fig2} we present the oscillatory regime of both fields after inflation in terms of cosmic time for the initial conditions $\phi_0 = 5.7M_p$ and $h_0 \approx 0.096M_p$. However, according to the argument we have presented above, during inflation the isocurvature effects are absent and the dynamics inside the valleys are an effective single-field description. Then the scalaron field should fulfill the single-field equation \cite{kofman1997, Durrer2022}

\beq 
    \ddot{\phi} + 3H\dot{\phi} + \dfrac{\dif V_\mathrm{eff}}{\dif \phi} \simeq 0\,,
    \label{SR EOM}
\enq 

and the evolution of the Higgs field is determined by the condition \eqref{higgs field condition Einstein}. It is worth mentioning that the effective single-field approximation is only valid during inflation within one of the valleys of the potential \eqref{potential Einstein frame}. The effective single-field inflation ends when the slow roll parameter $\epsilon$ is $\epsilon(t_\mathrm{end}) = 1$, which happens at the final value $\phi(t_\mathrm{end)} \approx 0.92 M_p$, where
\beq 
    t_\mathrm{end} = \dfrac{3}{4}\,H_0^{-1}\,e^{\alpha\phi_0}\,.
\enq 

After inflation, during the reheating era, the scalaron $\phi$ and the Higgs field $h$ oscillate around the global minimum $(0,0)$ of \eqref{potential Einstein frame}. That is, to follow the complete evolution of the fields after inflation, the system of equations \eqref{scalaron equation}, \eqref{higgs equation}, \eqref{1 Friedmann}, and \eqref{2 Friedmann} must be solved numerically with the appropriate initial conditions. Inflation within the valleys occurs for super-Planckian values of the field $\phi \gg M_p$ in order to produce sufficient inflation, i.e., $N_\mathrm{end} \gtrsim 60$. This condition is satisfied if $\phi_0 \simeq 5.7 M_p$, while the value of the Higgs field at the beginning of inflation is obtained by evaluating \eqref{higgs field condition Einstein} at $\phi_0$.

\section{Scalar fields in the Einstein frame}

In this section, we study spectator fields present in the Higgs-$R^2$ model such as free scalar fields, initially assuming no direct coupling between the matter fields and the inflationary sector in the Jordan frame. In other words, we get started in the Jordan frame and after conformal transformation, the matter sector is gravitationally coupled to the inflationary sector. This aims to induce direct couplings between the background composed of the fields $\phi, h$ and the scale factor $a(t)$, while also making it easier to solve the system’s dynamics. In the Jordan frame our model is given by

\beq
 \nonumber S_J  = \int \dif x^4 \sqrt{-g_J}\left[\dfrac{1}{2}(M_p^2 + \xi_h H^2)R_J +\dfrac{\xi_s}{4}\,R_J^2-\dfrac{1}{2}g^{\mu\nu}_J\,\partial_\mu h\,\partial_\nu h -\dfrac{\lambda}{4}\,h^4 \right]+S_\mathrm{matt}[\psi_i]\,\\
\enq 

where $\psi_i$ are the different matter fields that are present during and after inflation. In this work, we will consider a massive spectator scalar field $\chi$ that only interacts gravitationally. To the reader interested in the evolution and production of gauge bosons and other kind of matter fields after inflation, we recommend \cite{BEZRUKOV2019657} and the references therein. The action in the Jordan frame of a free scalar field $\chi$ is given by

\beq
   S_J[\chi] = \int \dif x^4\sqrt{-g_J}\left(-\dfrac{1}{2}g^{\mu\nu}_J\partial_\mu\chi\partial_\nu\chi-\dfrac{m_\chi^2}{2}\chi^2 - \dfrac{1}{2}\,\xi_\chi\,R_J\,\chi^2\right)\,.
    \label{boson action}
\enq 

 The dimensionless parameter $\xi_\chi$ represents the non-minimal coupling to the gravity sector and may take any value; however, studies based on electroweak vacuum stability \cite{markkanen2018} suggest that the value of non-minimal coupling constant should not exceed $10$. Moreover there are two interesting special cases: $\xi_\chi = 0$ and $\xi_\chi = 1/6$. The value $\xi_\chi =0$ is called the case of minimal coupling while $\xi_\chi = 1/6$ corresponds to conformal coupling. The first case corresponds to a free scalar field theory and the second one ensures the conformal invariance of the model if, in addition, $m_\chi = 0$.
 
 Although there is no direct coupling between matter fields and the inflationary sector in the first case, both scalaron $\phi$ and Higgs field $h$ gravitationally interact with every matter field present during inflation and after inflation, and a gravitational particle production process may occur. In the second one, free scalar $\chi$ is coupled to the Ricci scalar and the particle production may also occur as long as the matter fields are not conformally invariant, namely, if the scalar field is massive $m_\chi \neq 0$. These kinds of models, in which the scalar fields interact only gravitationally, are especially interesting. They can explain the particle production process, with the gravitational field responsible for creating $\chi$ particles. This process of particle creation due to gravitational interactions has implications for the reheating process \cite{dorsch2024, Markkanen2017, Ema_2017}, and it also plays a role in gravitational dark matter production \cite{Garcia2023, Garcia20231, Verner2023, Hashiba2019, Chung2001,ema2018production,long2023}. However, as pointed out in \cite{Watanabe2015}, the analysis of matter fields with non-minimal coupling in multi-field models of inflation becomes complicated even if the field metric is flat, as moving from one conformal frame to another induces non-canonical kinetic terms. The interpretation of the reheating phase after inflation also changes. While in the Jordan framework it is mainly due to the oscillatory behavior of the Hubble parameter, in the Einstein framework it is mainly due to the explicit interaction terms between the matter fields, and the inflationary sector of the model, through which particle production and reheating of the universe occurs.  Despite these differences, evaluating the inflationary dynamics is much simpler in the Einstein frame \cite{Kaiser2010}, as in this frame, Einstein’s equations are second-order, making their evaluation more straightforward \cite{faraoni1999, Postma2014}. Since the computation of the $\beta$-Bogolyubov coefficient is independent of the conformal frame \cite{Parker1969}, we choose to perform the calculation in the Einstein frame to avoid unnecessary complications when studying the background dynamics in the Jordan frame. While gravitational particle production is typically carried out in the Jordan frame \cite{Ema2016, Ema2017a, ema2018production, Ford2021}, performing the analysis in the Einstein frame \cite{Watanabe2015, belfiglio2024, GHOSHAL2025} is perfectly valid and yields similar results. 

As we assume that the energy density of the $\chi$ field never dominates, the reheating phase after inflation cannot proceed purely through gravitational interactions. Instead, it is assumed that reheating occurs via the decay of the scalaron into light particles $m_\chi \ll M$ \cite{vilenkin1985, Gorbunov2012, Li2021} and through the production of gauge bosons due to Higgs boson decay \cite{Bezrukov2009, BEZRUKOV2019657, He2021}. Thus, the spectator field $\chi$ is considered as dark matter produced gravitationally.
Given that the $\chi$-field interacts only gravitationally with the background fields $\phi$ and $h$, its dynamics can be analyzed effectively within the Einstein frame. To study its evolution, we apply the same conformal transformation of the metric \eqref{conformal transformation} as in the previous section. By doing so, we avoid considering back-reaction effects on the space-time geometry and the background dynamics. In the Einstein frame, we obtain the following action:

 \begin{multline}
    S_E[\tilde{\chi}] = -\dfrac{1}{2}\int\dif^4x\,\sqrt{-g}\left(g^{\mu\nu}\partial_\mu \tilde{\chi}\partial_\nu\tilde{\chi} + \xi_\chi R\tilde{\chi}^2\right.\\[0.3cm]
     +\left. e^{-\alpha\phi}m_\chi^2\tilde{\chi}^2\right) +3\alpha\left(\xi_\chi - \dfrac{1}{6}\right) \int\dif^4x\,\sqrt{-g}\left[ \tilde{\chi}\,g^{\mu\nu}\partial_\mu\tilde{\chi}\partial_\nu\phi\right.\\[0.3cm]
     \left. +\dfrac{\alpha}{4}\tilde{\chi}^2 g^{\mu\nu}\partial_\mu\phi\partial_\nu\phi \right]\,,
     \label{matter action Einstein}
 \end{multline}

where $\chi \to \tilde{\chi} = \Psi^{-1}\chi$ is the re-scaled scalar field and the Ricci scalar in the Einstein frame is given by $R = 6(\dot{H} + 2H^2)$. We can perform another conformal transformation which involves the scalar field $\tilde{\chi}$ leading to another Einstein frame to remove the non-minimal coupling constant $\xi_\chi$. However, we are only interested in the situation where $\chi$ is a spectator field and both scalaron field $\phi$ and Higgs field $h$ are responsible for the dynamics during and after inflation. One can see that the conformal transformation \eqref{conformal transformation} induces interaction terms between the scalaron $\phi$, the Higgs field $h$, and the free scalar $\tilde{\chi}$.   We can identify the interaction terms from the following action

\beq
S_\mathrm{int} = 3\alpha\left(\xi_\chi - \dfrac{1}{6}\right) \int\dif^4x\,\sqrt{-g}\left[ \tilde{\chi}\,g^{\mu\nu}\partial_\mu\tilde{\chi}\partial_\nu\phi  +\dfrac{\alpha}{4}\tilde{\chi}^2 g^{\mu\nu}\partial_\mu\phi\,\partial_\nu\phi \right]\,.
     \label{interaction action}
\enq 

This action describes the interaction between the background fields $\phi$ and $h$ with the massive spectator field $\tilde{\chi}$. We can see that even with $\xi_\chi = 0$, this interaction does not disappear unless the field $\tilde{\chi}$ is conformally invariant ($m_\chi = 0\quad \xi_\chi = 1/6$). Since we are considering $\chi$ as a spectator dark matter field that interacts only gravitationally, it is natural to explore its production mechanisms within inflationary models. In the literature, there are studies on the production of dark matter in the $R^2$ model \cite{Gorbunov2012, Bernal2020, Li2021, racco2024}, where a scalar field $\chi$ with action \eqref{boson action} is considered. In  \cite{Gorbunov2012, Bernal2020, Li2021}, the authors study the gravitational production of dark matter in the $R^2$ inflation model and show that the scalaron $\phi$ can decay perturbatively into dark matter particles $\chi$ with masses $m_\chi \ll M$, where $M \simeq 1.3\times 10^{-5}M_p$ is the mass term of the scalaron. The corresponding two-body decay width is given by \cite{vilenkin1985, Gorbunov2012, Bernal2020}.

\beq 
\Gamma_{\phi\to \chi\chi} = \dfrac{1}{192\pi M_p^2}\left(1 - 6\xi_\chi + 2\dfrac{m_\chi^2}{M^2}\right)^2\sqrt{1 - \dfrac{4 m_\chi^2}{M^2}}
\enq

One can see that for heavy particles $m_\chi \gtrsim M$, perturbative particle production is kinematically forbidden. The production of light particles after inflation occurs mainly through this mechanism \cite{Gorbunov2012}. However, in \cite{racco2024} the authors study the gravitational production fo heavy dark matter particles with minimal coupling during and after inflation. They find that no additional particle production occurs during inflation if the particle mass is smaller than the Hubble scale $H_\mathrm{inf}$ during inflation. However, if the mass is larger, the production after inflation becomes dominant.
On the other hand, there have been few attempts to study spectator dark matter field in the Higgs-$R^2$ model. In \cite{Samart2019}, the authors find that the Higgs-$R^2$ model is equivalent to a scalar singlet dark matter or Higgs-portal model, constraining the scalar singlet dark matter sector using the dark matter abundance and model parameters. This work, however, does not consider an additional scalar field coupled to the background. In \cite{aoki2022reheating}, the authors introduce an extra scalar degree of freedom with non-minimal coupling that plays the role of freeze-in dark matter. They also discuss the freeze-in production of dark matter both from the non-thermal scattering during reheating and the thermal scattering after reheating. 

In our case, we investigate the gravitational production of the $\chi$ field through non-perturbative mechanisms, taking into account the induced interaction terms in the Einstein frame. Specifically, particle production can occur due to the expansion of the universe and the dynamics of the fields involved during the inflationary epoch.
The classical time-dependent background in this scenario is determined by the evolution of the scale factor $a(t)$ and the scalar fields $\phi$ and $h$. That is, the spectator field $\chi$, which is initially decoupled from other matter fields, propagates in a time-dependent classical background governed by the evolution of $\phi$ and h in the Higgs-$R^2$ model.

To analyze this process, we need to derive the equation of motion for the spectator field $\tilde{\chi}$ and track the evolution of its modes during and after inflation. By varying the action \eqref{matter action Einstein} with respect to $\tilde{\chi}$, we obtain the following equation of motion:

\beq
 (\square -e^{-\alpha\phi}\,m_\chi^2 - \xi_\chi\,R)\tilde{\chi}
    + \left[\dfrac{\alpha}{2}\,\left(1-6\xi_\chi\right)\square\phi -\dfrac{\alpha^2}{4}\left(1-6\xi_\chi\right)(\partial\phi)^2 \right]\tilde{\chi} = 0\,.
    \label{EOM}
\enq

 Notice that the background dynamics is explicit in the equation of the spectator field $\tilde{\chi}$ due to the conformal transformation. This induces a time-dependent effective mass term, $m_\mathrm{eff}^2(t),$ for the spectator field $\tilde{\chi}$. Consequently, in the Einstein frame, the model presents an interaction term with the dynamics of the classical background through the Ricci scalar, the scale factor $a$, and the dynamics of the fields $\phi$ and $h$. As a result, studying the solutions is generally highly complex from an analytical perspective; hence a numerical treatment of the background dynamics is necessary for a comprehensive analysis of mode evolution, especially after inflation when both fields $\phi$ and $h$  exhibit a strong interacting anharmonic oscillatory behavior, as one can see in the Fig. \ref{fig2}. By numerically determining the background evolution of the scalaron, the Higgs field, and the scale factor, we can compute the Ricci scalar using $R = 6(\dot{H} + 2H^2)$,  ensuring that the dynamics are well-determined. However, during effective inflation along one of the potential's valleys, we can approximate the background dynamics using the single-field equation of motion \eqref{SR EOM}.

\section{Mode evolution}

In this section, we numerically solve the equation of motion \eqref{EOM} along with the background equations \eqref{scalaron equation}, \eqref{higgs equation}, \eqref{FLRW metric} to obtain the mode functions $\chi$ for calculating the number density of particles produced, $n_k$. The equation of motion for $\tilde{\chi}$ considering a FLRW universe \eqref{FLRW metric} can be written in terms of the cosmic time $t$ as

\beq 
\ddot{\tilde{\chi}} + 3H\dot{\tilde{\chi}} + \Omega^2(t) \tilde{\chi} =0\,,
\enq

where $\Omega^2(t)$ is a background variables dependent and is given by 

\beq 
  \Omega^2(t) = e^{-\alpha\phi}m_\chi^2 + \dfrac{k^2}{a^2} + \xi_\chi R
    + 3\alpha\left(\dfrac{1}{6} - \xi_\chi \right)(\ddot{\phi} + 3H\dot{\phi}) - \dfrac{3\alpha^2}{2}\left(\dfrac{1}{6} - \xi_\chi \right)\dot{\phi}^2\,.
\enq 

Rescaling the field $\tilde{\chi} \to a^{-3/2}(t)\,\tilde{\chi}(t,\mathbf{r})$ and using the background equations of motion \eqref{scalaron equation}, \eqref{2 Friedmann} the equation for the scalar field $\tilde{\chi}$ becomes a oscillator-like equation with time-dependent frequency 

\beq 
\ddot{X}(t,\mathbf{r}) + \omega^2(t)\,X(t,\mathbf{r}) = 0\,,
\enq

where $\omega^2(t)$ is given by 

\beq 
\omega^2(t) = \Omega^2(t) - \dfrac{9}{4}H^2 - \dfrac{3}{2}\dot{H}\,.
\enq 

Particle production in the early universe is a process of quantum field theory in curved spacetime, where the gravitational field is not quantized, but the free scalar field $\chi$ is quantized. Therefore, the production of $\chi$ particles is a quantum effect \cite{Parker1969, Ford1987, kolb2023}. We proceed by quantizing the spectator field in the Heisenberg picture so that the $\chi$ field and its conjugate momentum $\pi_\chi = \dot{\chi}$ are treated as field operators. In Fourier space, the rescaled field can be written as 

\beq
    \nonumber X(t,\mathbf{r}) = \int \dfrac{\dif^3 \mathrm{k}}{(2\pi)^3}\left(\hat{a}_\mathbf{k}\,X_k(t)\,e^{-i\mathbf{k}\cdot\mathbf{r}} + \hat{a}^\dagger_\mathbf{k}\,X^*_k(t)\,e^{i\mathbf{k}\cdot\mathbf{r}}   \right)\,,\\
\enq 

where $\mathbf{k}$ is the comoving wave number and $\hat{a}_\mathbf{k}$, $\hat{a}^\dagger_\mathbf{k}$ are the creation and annihilation operators respectively, which obey the standard commutation relations at equal times $[a_\mathbf{k}\,,a^\dagger_\mathbf{k'}]= (2\pi)^3\delta^{(3)}(\mathbf{k}-\mathbf{k}')$ and $[a_\mathbf{k}\,,a_\mathbf{k'}] =[a^\dagger_\mathbf{k}\,,a^\dagger_\mathbf{k'}] =0$. The field equation for Fourier mode $X_k(t)$ is given by 

\beq 
\ddot{X}_k(t) + \omega_k^2(t)X_k(t) = 0\,,
 \label{mode equation}
\enq

with

\beq
    \omega_k^2(t) = \left(\dfrac{k}{a}\right)^2 + m_\mathrm{eff}^2(t)\,,
    \label{frequency}
\enq 

and the following effective mass term 

\beq 
 m_\mathrm{eff}^2(t) =  e^{-\alpha\phi}m_\chi^2 + \xi_\chi\,R - \dfrac{9}{4}H^2 - \dfrac{3}{2}\dot{H}
    +3\alpha\left(\xi_\chi - \dfrac{1}{6}\right)(V_\phi - \alpha M_p^2 \dot{H})\,.
    \label{mass term infalation}
\enq

The equation for $X_k$ describes an oscillator with time-dependent frequency $\omega_k^2(t)$, hence to track the evolution of the modes $X_k$ during and after inflation, it is necessary to analyze $\omega_k^2(t)$. One can see that during the quasi De Sitter phase during inflation when the effective single field inflation is valid, the effective mass is simplified because of $\dot{H}\to 0$. In this case, one can see that the effective mass takes the form

\beq
  m_\mathrm{eff}^2(t \ll t_\mathrm{end}) \simeq e^{-\alpha\phi} m_\chi^2 + 12 \left(\xi_\chi - \dfrac{3}{16}\right) H^2
    + 3\alpha\left(\xi_\chi - \dfrac{1}{6}\right)V_\phi\,.
    \label{effective mass}
\enq

\begin{figure}[h]
\centering
\includegraphics[width=0.8\textwidth]{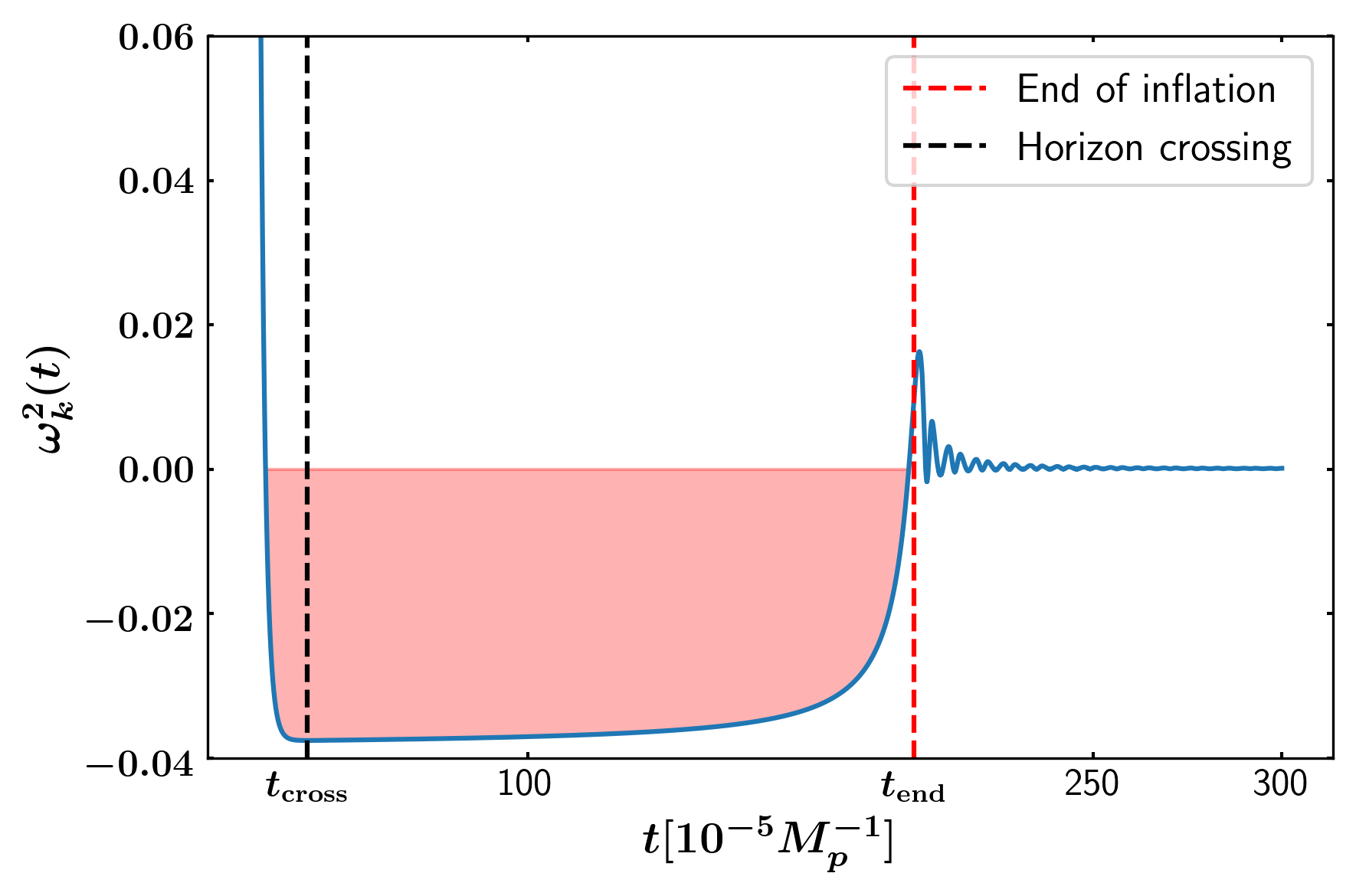}
\caption{Time-dependent frequency as a function of cosmic time t for a mass $m_\chi = 10^{-2}H_\mathrm{end}$, a non-minimal coupling $\xi_\chi = 0$, and $k \gg aH$ for 5 e-folds before horizon crossing. We set $N_e = 60$ before the end of inflation as the moment when the mode crosses the horizon. }
   \label{fig3}
\end{figure}

 We can see that the mass term is exponentially suppressed during inflation when $\phi \gg M_p$, making the mass of the spectator field much smaller than the Hubble parameter at the end of inflation, $m_\chi \ll H_\mathrm{end}$. However, after inflation, when the scalaron takes values $\phi \ll M_p$, the free scalar $\tilde{\chi}$ becomes super heavy, $m_\chi \gg H_\mathrm{end}$, due to the exponential dependence on the scalaron $\phi$. Then at the beginning of inflation, when $t \ll t_\mathrm{end}$ the Ricci scalar is $R \simeq 12H^2$ and the dominant terms in \eqref{frequency} are $k^2/a^2$, $H^2$ and $12\xi_\chi H^2$. However, towards the end of inflation, the term $\dot{H}$ cannot be neglected, as after inflation, $\dot{H}$ and the background oscillations dominate the dynamics. The term $V_\phi$ regulates the oscillations of the effective mass after inflation, but it is small compared to $H^2$ and $\dot{H}$. The contribution of each term depends on the moment at which it is considered. The mass term benefits from exponential growth when $\phi \ll M_p$, as we anticipated earlier. Therefore, after inflation, the dominant terms are the mass term and $\dot{H}$. Additionally, if the coupling constant is $\xi_\chi < 1/6$, the effective mass could be negative during inflation. In such a case, the frequency $\omega_k^2(t)$ can become negative for super-horizon modes $k \ll aH$, leading to an exponential mode growth $\tilde{\chi} \sim e^{|\omega_k|t}$ during inflation due to tachyonic instability. This regime may result in a very efficient amplification of particle production \cite{Markkanen2017,dufaux2006preheating}. However, for large enough modes $k$, namely

\beq 
k^2 > k_*^2 = -m_\mathrm{eff}^2(t)\,,
 \enq

have positive frequency $\omega_k^2(t)> 0$ even if $m_\mathrm{eff}^2 < 0$, which corresponds to sub-horizon regime $k\gg aH$. This occurs when the modes are well inside the horizon at the
beginning of inflation. In the Fig. \ref{fig3} we show the evolution of $\omega_k^2(t)$ for a non-minimal coupling $\xi =0$. Before $t_\mathrm{cross}$, the frequency is dominated by the decay of the physical mode $k/a \sim e^{-Ht}$. Once the mode crosses the horizon $(k \ll aH)$, $\omega_k^2(t)$ remains almost constant, and its value is negative since the coefficient of the dominant term $H^2$ is negative. For $t \gtrsim t_\mathrm{end}$, the frequency increases and oscillates after inflation as the background terms oscillate around the minimum of the potential (preheating).

One can also observe growth due to the exponential dependence of the scalaron in the mass term toward the end of inflation. To numerically solve equation \eqref{mode equation} along with the background equations, we must specify suitable initial conditions that ensure the modes are within the horizon at early times. In the asymptotic limit $t \to 0$ and according to the behavior of $\omega_k^2(t)$ at the beginning of inflation, the dominant term is $k/a$. In this limit, the modes are sub-horizon $k \gg aH$, and the modes exhibit Minkowski-like behavior. This motivates choosing the Bunch-Davies vacuum as the initial condition \cite{kofman1997}

\beq
\lim_{t\to 0} X_k(t) = \dfrac{1}{\sqrt{2\omega_k}}\,e^{-i\omega_k t}\,.
\label{BD vacuum}
\enq 

As previously mentioned, for certain values of $\xi_\chi$, the modes experience tachyonic excitations during the time they are outside the horizon. Therefore, to ensure that each mode is within the horizon and to use \(\eqref{BD vacuum}\) as the initial condition, we must track the evolution of the corresponding mode at times $t_\mathrm{ini}$ when the mode is inside the horizon. 
For concreteness, we assume that the Planck pivot scale $k_*$ leaves the horizon $N_* = 60$ e-folds before the end of inflation with a total duration of inflation of $N_\mathrm{tot} = 76.150$ e-folds. We also take the initial conditions to be the values of $X_k$ and the background variables 5 e-folds before the mode crosses the horizon. This guarantees that the initial conditions for the modes $X_k(t)$ and their derivatives $\dot{X}_k(t)$ can safely be set to the Bunch-Davies type.

\section{Description of particle production}

Once we have established the behavior of mode functions $X_k$ during and after inflation for some values of spectator mass, the non-minimal coupling $\xi_\chi$ and comoving wavenumber $k$, it turns to analyze the non-perturbative particle production. The production of $\chi$ particles is due to the time dependence of the background variables, in particular, the non-adiabatic behavior of the frequency $\omega_k(t)$ that induces a mix among negative- and positive-frequency \cite{Parker1969}. The adiabaticity condition is  given by the adiabatic parameter $A_k$ defined as

\begin{figure}[h]
\centering
\includegraphics[width=0.8\textwidth]{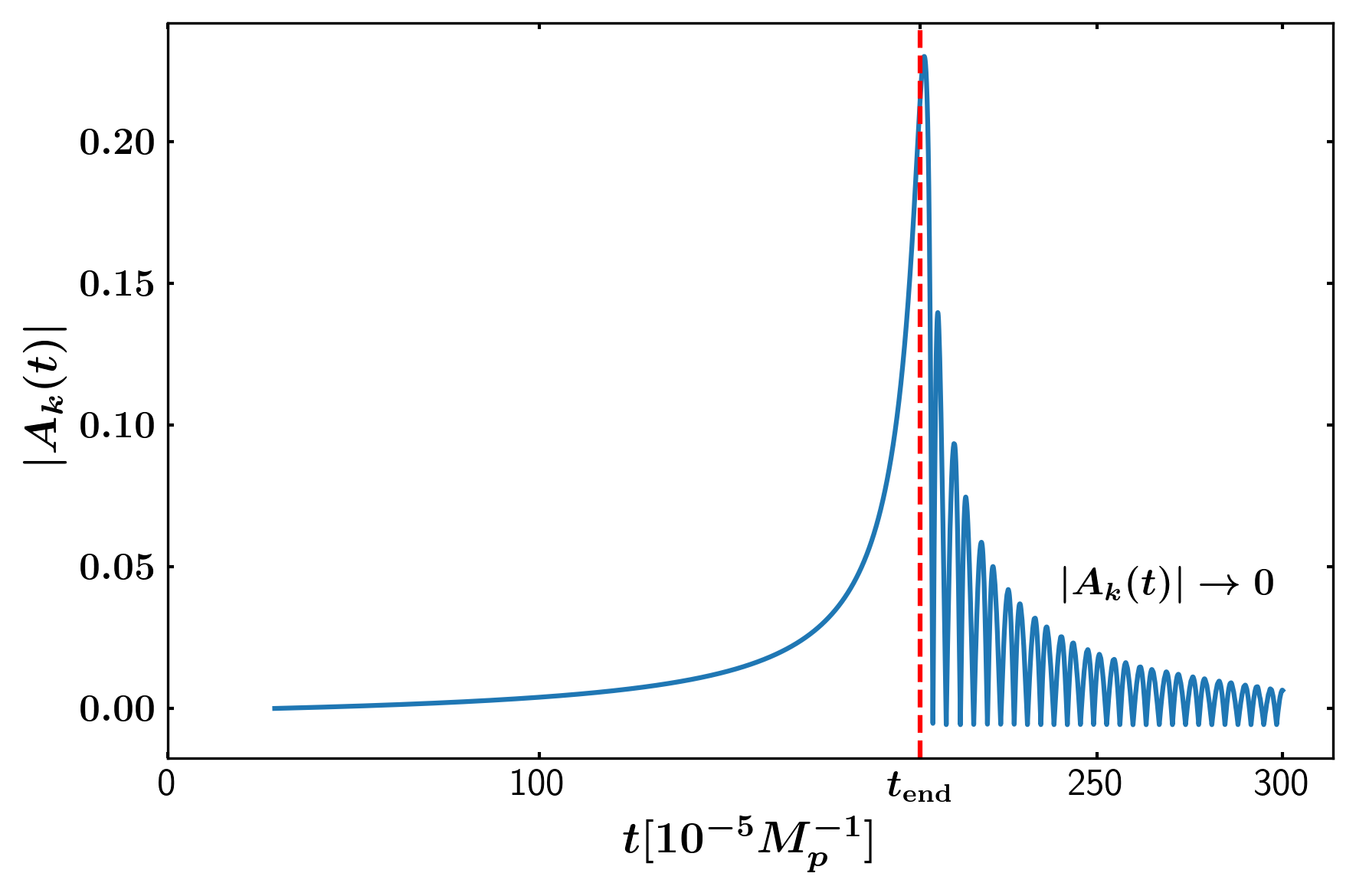}
\caption{An example of non-adiabatic mode evolution leading the gravitational particle production of $\chi$ particles described by the adiabaticity parameter $|A_k(t)|$ for $m_\chi = 10^{-2}H_\mathrm{end}$ and $\xi_\chi = 1/6$. At the end of inflation, one sees an abrupt increase on $|A_k(t)|$ and non-adiabaticity is induced, leading to gravitational particle production $n_k(t)\neq 0$. At late times, when the stabilization occurs, we recover the adiabaticity evolution. }
   \label{fig4}
\end{figure}

\beq 
A_k(t) = \dfrac{\dot{\omega}_k}{\omega_k^2}\,.
\enq

The adiabatic evolution of $\omega_k(t)$ under the WKB approximation of the modes $X_k(t)$ guarantees that the positive-frequency mode at early times evolves into positive-frequency modes at late times. That is, for asymptotically early times $t\to 0$, the modes $X_k(t)$ evolve as \cite{winitzki2005cosmological}

\beq
\lim_{t\to 0} X_k(t) = \dfrac{e^{-i\int^t \omega_k(t')\dif t'}}{\sqrt{2\omega_k(t)}}\,.
\label{WKB mode}
\enq

This behavior usually happens at early times $t_i\to 0$ and at sufficiently late times $t_f\to +\infty$, but for intermediate times $t_i<t_*<t_f$ during the early universe, the evolution of the modes is no longer adiabatic  $|A_k|\gg 1$. During inflation, the positive-frequency solution corresponds to the Bunch-Davies vacuum \eqref{BD vacuum}, as described in the previous section. This evolution remains valid as long as the parameter $|A_k| \ll 1$ at all times. While this condition is met, no particle production occurs, as adiabatic evolution is ensured, and there is no mixing of negative- and positive-frequency modes. From Fig. \ref{fig3}, we can see that the evolution of $\omega_k(t)$ is smooth until the end of inflation. At that point, the condition $|A_k| \ll 1$ is strongly violated, leading to an enhancement in particle production, and gravitational particle production is enhanced at the time $t_*$, corresponding to the end of inflation $t_\mathrm{end}$. We show the non-adiabaticity evolution in Fig. \ref{fig4} parametrized by the adiabatic parameter $|A_k(t)|$.  The mixture of positive and negative frequency modes at time $t_\mathrm{end}$ can be represented as the sum of the solutions $\exp(\pm i\int^t \omega_k(t')\dif t')$ using the WKB approximation \cite{kofman1997, kolb2023}

\beq 
 X_k(t) = \dfrac{\alpha_k(t)}{\sqrt{2\omega_k(t)}}\,e^{-i\int^t \omega_k(t')\dif t'} + \dfrac{\beta_k(t)}{\sqrt{2\omega_k(t)}}\,e^{i\int^t \omega_k(t')\dif t'}\,,
 \label{WKB}
\enq 

where $\alpha_k(t)$ and $\beta_k(t)$ are the time-dependent Bogolyubov coefficients. Comparing with the Bunch-Davies initial condition \eqref{WKB mode}, one see thats $\alpha_k(t_0) = 1$ and $\beta_k(t_0) = 0$. 
Gravitational particle production is a quantum process that occurs in the early universe, primarily near the end of inflation and continues for a few e-folds afterwards until the $\beta$-Bogolyubov coefficient is stabilized, allowing for a well-defined computation of the energy density of the produced particles. However, if the initial and final vacuum states are not the same, particle production can also be interpreted in terms of asymptotic in and out states, highlighting its connection to both early and late-time behavior. We define the initial vacuum state with the corresponding creation-annihilation operators $\hat{a}^\dagger_\mathbf{k}$ and $\hat{a}_\mathbf{k}$ by 

\beq 
\hat{a}_\mathbf{k}\left|0\right>_\mathrm{early} = 0\,.
\enq 

The final state will have a different set of creationist-annihilation operators, and thus a different vacuum state. $\hat{a}_\mathbf{k}$ it is related to the late-time operators $\hat{b}^\dagger_\mathbf{k}$ and $\hat{b}_\mathbf{k}$ by a Bogolyubov tranformation

\beq 
\hat{a}_\mathbf{k} = \alpha^*_\mathrm{k} \hat{b}_\mathbf{k} - \beta^*_\mathrm{k}\hat{b}^\dagger_\mathrm{k}\,,
\enq 

where $\alpha_\mathbf{k}$ and $\beta_\mathbf{k}$ are the time-independent Bogolyubov coefficients. The comoving particle number density $n_\chi(t)$ is related to the coefficient $\beta_k$ of the Bogolyubov transformations which relate the vacuum states at the asymptotic limits $t\to 0$ and $t\to \infty$ and coincides with $\beta_k(t)$ at late times. We can compute $\beta_k(t)$ by solving \eqref{mode equation} and evaluating

\beq
|\beta_k(t)|^2 = \dfrac{\omega_k}{2}|X_k|^2 + \dfrac{|\dot{X}_k|^2}{2\omega_k} - \dfrac{1}{2}\,.
\label{beta}
\enq 

The total number comoving density $n_\chi(t)$ of created particles in the late-time limit $t\to\infty$ is given by

\beq 
 a^3(t)\,n_\chi(t) = \int_0^\infty \dfrac{\dif^3 k }{(2\pi)^3}|\beta_k(t)|^2\,,
\label{particle number}
\enq 

which can be rewritten as 

\beq 
a^3(t)n_\chi(t) = \int \dfrac{\dif k}{k} n_k(t)\,,\quad n_k(t) = \dfrac{k^3}{2\pi^2}|\beta_k(t)|^2
\enq 

where $n_k(t) $ is the comoving spectral number density of particles. Notice that the expression \eqref{beta} is only valid when $\omega_k^2(t) >0$, so the comoving number density of particles created is valid at times when the frequency \eqref{frequency} is positive. 
In some cases, it is possible to analytically evaluate the $\beta$-Bogolyubov coefficient. The expression \eqref{beta} is typically used in numerical approaches, as, in most cases, evaluating $\beta_k$ numerically requires knowledge of the mode dynamics $X_k(t)$ and the frequency $\omega_k(t)$. That is, we need to numerically solve the equation \eqref{mode equation} with Bunch-Davies initial conditions. However, there is an alternative method to compute the Bogolyubov coefficients $\alpha_k$ and $\beta_k$. The WKB approximation \eqref{WKB} provides a solution to \eqref{mode equation} if the time-dependent Bogolyubov coefficients $\alpha_k(t)$ and $\beta_k(t)$ satisfy the system of equations given in \cite{kofman1997}.

\beq 
\dot{\alpha}_k = \dfrac{\dot{\omega}_k}{2\omega_k} e^{2i\psi_k(t)}\,\beta_k\,,\quad \dot{\beta}_k = \dfrac{\dot{\omega}_k}{2\omega_k} e^{-2i\psi_k(t)}\,\alpha_k\,,
\enq 

where $\psi_k(t) = \int_{t_i}^t \dif t' \, \omega_k(t')$ is a time-dependent oscillatory phase. Gravitational particle production occurs in the Bunch-Davies regime, where $|\beta_k| \ll 1$ and $|\alpha_k| \simeq 1$. In this case, the Bogolyubov coefficient $\beta_k$ is given by

\beq 
|\beta_k(t)| \simeq \dfrac{1}{2}\int_{t_i}^t \dif t'\, \dfrac{\dot{\omega}_k}{\omega_k} \exp\left(- 2i \int_{t_i}^{t'}\dif t'' \omega_k(t'')  \right)\,.
\label{integral beta}
\enq 

It is possible to numerically solve this integral without the need to numerically solve the equation \eqref{mode equation} for an initial time $t_i \ll t$; however, the numerical calculation of the integral becomes complex in most cases when the oscillatory phase $\psi_k(t)$ oscillates rapidly. This approach is more useful when one seeks to evaluate $\beta_k$ analytically or approximately. To do so, the steepest descent method is commonly employed in the literature \cite{Chung2003, Hashiba2019, racco2024, jenks2025}. We have chosen to adopt a numerical approach to compute \eqref{particle number}, due to the nature of the model, which involves a non-minimal coupling and a nontrivial potential term. These features significantly complicate the evaluation of the integral expression \eqref{integral beta}. As this analysis lies beyond the main scope of the present work, we leave it for future investigations.

\subsection{Numerical results}

What follows is to find the spectrum of $\chi$ particles and their evolution in the cosmic time $t$. Hence we numerically solve \eqref{mode equation}, constructed the Bogolyubov coefficient $|\beta_k(t)|$, and compute the spectrum $n_k$ using \eqref{particle number}. In the previous section, we described the procedure to numerically evaluate the set of equations to ensure that each mode $k$ is within the horizon at the beginning of the evolution, which allows us to use the Bunch-Davies vacuum \eqref{BD vacuum} as an initial condition of the mode equation for $X_k(t)$. The initial conditions for the background are set at time $t_\mathrm{ini}$, corresponding to the moment when the modes are within the horizon. For example, for $N_* = 60$ at the time of crossing, the initial time corresponds to $t_\mathrm{ini} \approx 28.712$ in units of $[10^{-5}M_p^{-1}]$, and the corresponding mode $k = aH$ will be the value of $aH$ at time $t_\mathrm{ini}$.  In this way, we can associate a value of $k$ to each value of $N_e$, which determines the scale $k$ that crosses the horizon $N_e$ e-folds before the end of inflation.

\begin{figure}[h]
\centering
\includegraphics[width=0.6\textwidth]{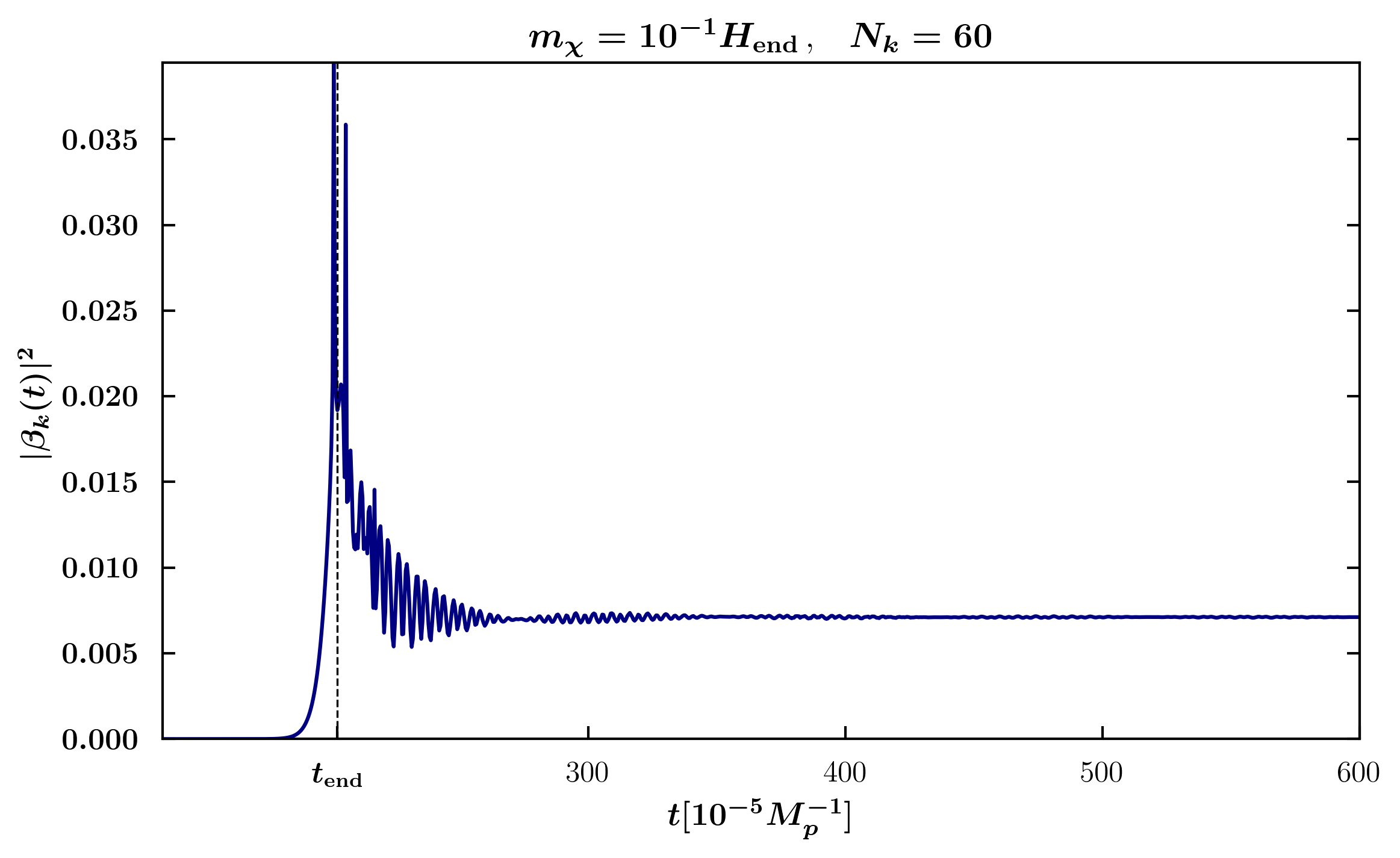}\\
\includegraphics[width=0.6\textwidth]{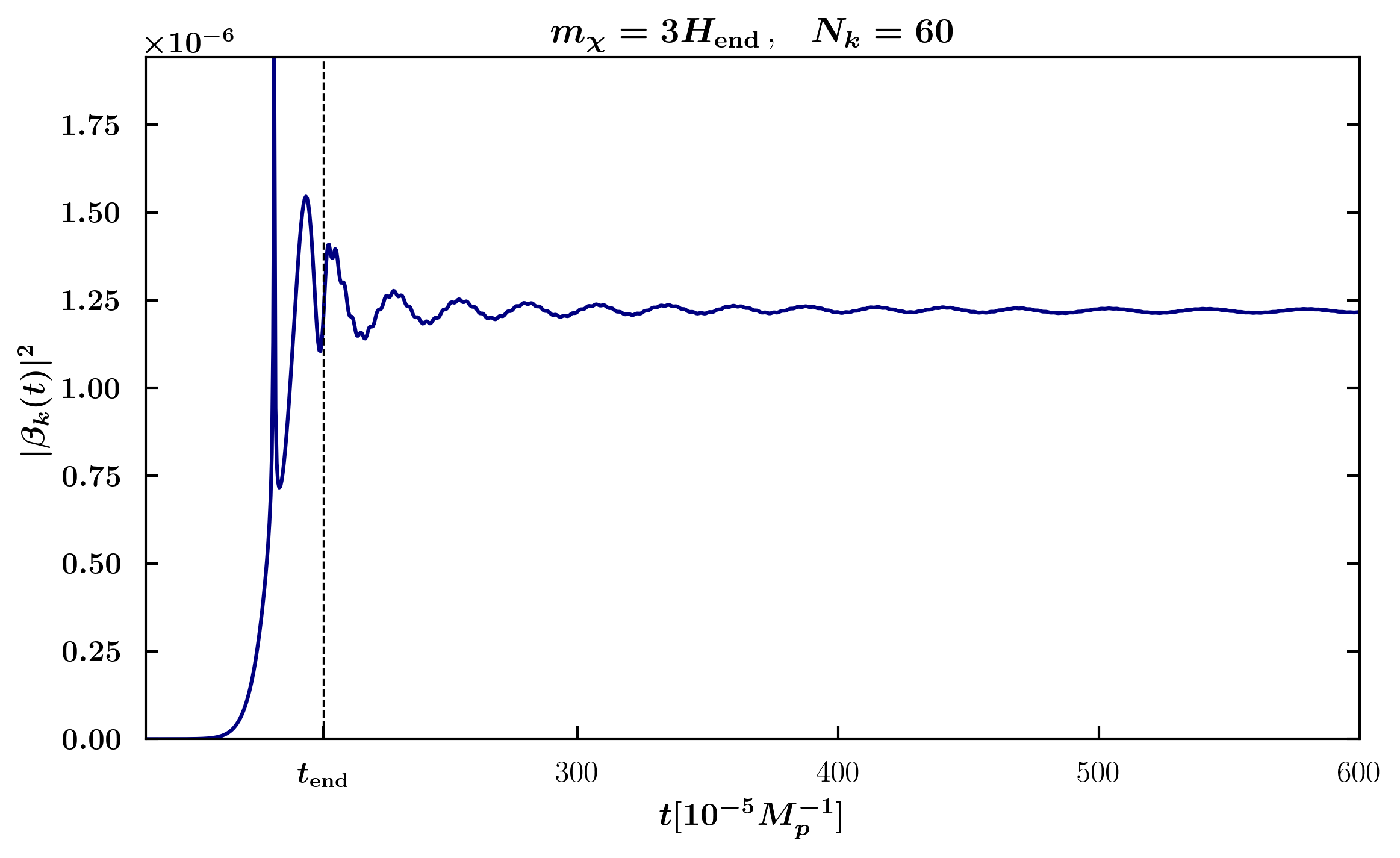}
\caption{The Bogolyubov coefficient $\beta_k$ is shown as a function of cosmic time $t$ for a spectator field with masses $m_\chi = 10^{-1}H_\mathrm{end}$ and $m_\chi = 3H_\mathrm{end}$, corresponding to a mode that exits the horizon $N_k = 60$ e-folds before the end of inflation. We observe that $\beta_k$ increases near the end of inflation and eventually stabilizes at late times, for $t \gg t_\mathrm{end}$.}
   \label{fig5}
\end{figure}

An alternative approach to selecting the $k$-modes is to work directly with the dimensionless ratio $k/k_{\mathrm{end}}$ in the equation for $X_k(t)$. In this framework, modes with $k < k_{\mathrm{end}}$ will have crossed the horizon at some point before the end of inflation. However, for our purposes, it is more convenient to generate a list of modes that satisfy $k = aH$ by solving the background equations and associating each mode with the corresponding number of e-folds $N_e$ before the end of inflation. The resulting list of physical modes $k$ is then normalized by $k_{\mathrm{end}}$ to construct the spectrum $n_k(t)$. The Bogolyubov coefficient $|\beta_k(t)|^2$ is computed using Eq.~\eqref{beta}, and the comoving number density of gravitationally produced particles is determined by $n_k(t)$. It is important to note that the particle number is evaluated only once it has stabilized; that is, we extract the value of $n_k$ at the end of the numerical simulation to ensure that it has reached a steady state.

\subsubsection{Conformal coupling $\xi_\chi = 1/6$}

We first present the results for conformally coupled particles ($\xi_\chi = 1/6$). As an example, we present the time evolution of the $\beta$-Bogolyubov coefficient as a function of time for two different cases: light particles $m_\chi < H_\mathrm{end}$ and heavy particles $m_\chi \gtrsim H_\mathrm{end}$. We should note that the expression for $\beta_k(t)$ is only valid when $\omega_k^2(t) >0 $ , we point out that $n_k(t)$ only makes physical sense at late times $t \to \infty$ \cite{kolb2023} when its value stabilizes. 

\begin{figure}[h]
\centering
\includegraphics[width= 0.89 \textwidth]{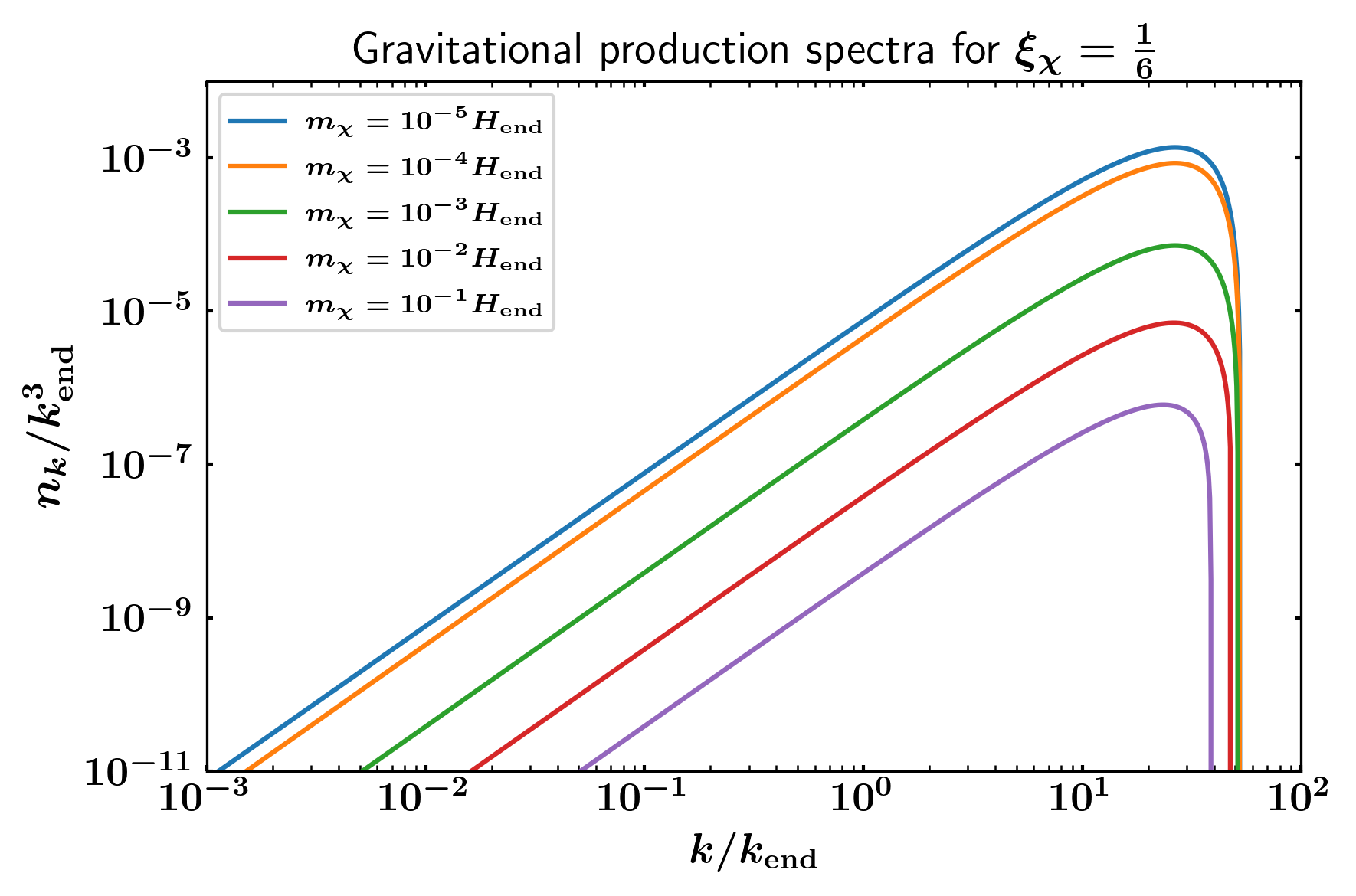}
\caption{Numerical $n_k/k_\mathrm{end}^3$ vs $k/k_\mathrm{end}$ spectra of conformally-coupled scalars for differente light masses $m_\chi$. One can see that the UV  tail is suppressed for this range of masses. }
   \label{fig6}
\end{figure}

The frequency \eqref{frequency} for scalar conformally coupled turns out to be 

\beq 
\omega_k^2(t) = \left(\dfrac{k}{a}\right)^2 + e^{-\alpha\phi}m_\chi^2 + \dfrac{R}{6} - \dfrac{9}{4}H^2 - \dfrac{3}{2}\dot{H}
\enq

which can be negative for $k\ll aH$. For this reason, we take into account the sub-horizon evolution of the modes $X_k(t)$ and we take $k$ deep inside the horizon at the beginning of the simulation.  We present the two different cases in the Fig. \ref{fig5}:  $m_\chi = 10^{-1} H_\mathrm{end}$ and $m_\chi = 3H_\mathrm{end}$. One can observe that the number of particles created saturates at the end of inflation, stabilizing and becoming constant at late times $t \to \infty$. This growth in $\beta_k(t)$ is due to the non-adiabaticity of $\omega_k$ at the end of inflation, as can be seen in Fig. \ref{fig4}. The oscillations of $\beta_k(t)$ after inflation arise from resonances occurring at specific moments between the oscillations of the modes $X_k$ and the frequency itself. During this epoch, particle production is driven primarily by the oscillatory behavior of the scalaron and the Higgs field, as illustrated in Fig. \ref{fig2}, through the term $\dot{H}$ and the Ricci scalar $R$ via the non-minimal coupling $\xi_\chi = 1/6$. 
For both cases, the particle number is well-defined, grows from the end of inflation, and stabilizes at late times, as expected. We can also observe that particle production is more efficient for lighter species, as indicated by the value of $|\beta_k(t)|^2$ once particle production has ceased. In particular, we find $|\beta_k(t_\mathrm{final})|^2 \approx 0.0071$ for the case $m_\chi = 10^{-1}H_\mathrm{end}$, while for $m_\chi = 3 H_\mathrm{end}$ we obtain $|\beta_k(t_\mathrm{final})|^2 \approx 1.216 \times 10^{-6}$. This behavior is precisely what one expects: heavier particles are produced less efficiently via gravitational mechanisms, as more energy is required to excite them from the vacuum.

\begin{figure}[h]
\centering
\includegraphics[width= 0.89 \textwidth]{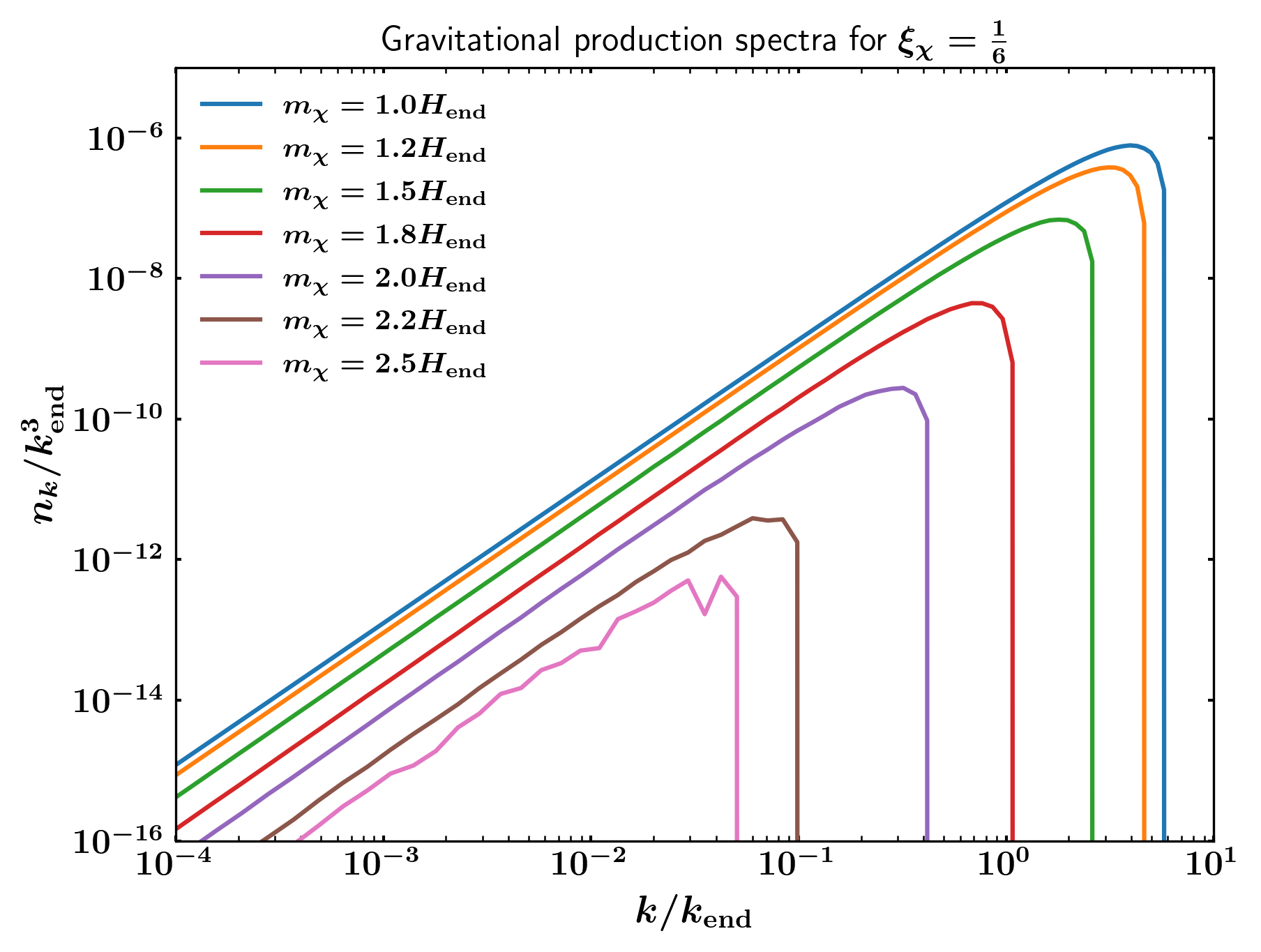}
\caption{Right: Comoving number density of particles as a function fo cosmic time $t$. Left: Numerical $n_k/k_\mathrm{end}^3$ vs $k/k_\mathrm{end}$ spectra of conformally-coupled scalars for differente large masses $m_\chi \gtrsim H_\mathrm{end}$. One can see that the UV  tail is suppressed for this range of masses. }
   \label{fig7}
\end{figure}

We have also numerically computed the spectra of the comoving number density $n_k(t)$ as a function of $k/k_\mathrm{end}$ for a range of light masses varying from $10^{-5}H_\mathrm{end}$ to $10^{-1}H_\mathrm{end}$. Our findings are summarized in Fig. \ref{fig6}. From the figure, one can observe several remarkable features.  Conformally coupled scalars show a spectra that are strongly peaked at some value $k_\mathrm{peak}$. 
For IR modes $(k \ll k_\mathrm{end})$, we see that the spectrum is blue-tilted, reaching a peak at some intermediate value $k_\mathrm{peak} > k_\mathrm{end}$, and then is suppressed for UV modes $(k \gg k_\mathrm{end})$ due to inflation, that is, those that remain within the horizon during inflation.  IR modes leave the horizon during inflation and re-enter later during the reheating epoch after inflation. Therefore, the dynamics of the $X_k(t)$ modes are dictated by the effective mass term \eqref{effective mass} with a quasi-De Sitter spacetime as background with $H\simeq \mathrm{constant}$. This results in a comoving number density with a scaling $n_k \sim k^3$ for these modes. That is, in the IR region, the particle occupation number vanishes $n_k \to 0$ as $k\to 0$, consistent with what is observed in the figure. On the other hand, UV modes correspond to those that never crossed the horizon during inflation, and therefore their production is strongly suppressed. In place, these modes are mainly produced  during the transition to the matter-dominated era and in general during the reheating era. This behavior is characteristic of gravitational particle production and has been extensively reported in the literature \cite{Herring2020, Kaneta2022, Garcia20231, Garcia2023, Verner2023}.

The field's mass $m_\chi$ plays a key role in determining the amplitude of $n_k$ at the peak momentum $k_\mathrm{peak}$, which systematically decreases as $m_\chi$ increases \cite{Garcia2023, Verner2023, long2023}.  For large masses $m_\chi \gtrsim H_\mathrm{end}$, the situation is quite similar. As the mass increases, the particle production tends to decrease due to the increase of the effective mass. This behavior can be clearly seen in the particle spectrum $n_k$ shown in Fig. \ref{fig7}. 

\begin{figure}[h]
\centering
\includegraphics[width= 0.89 \textwidth]{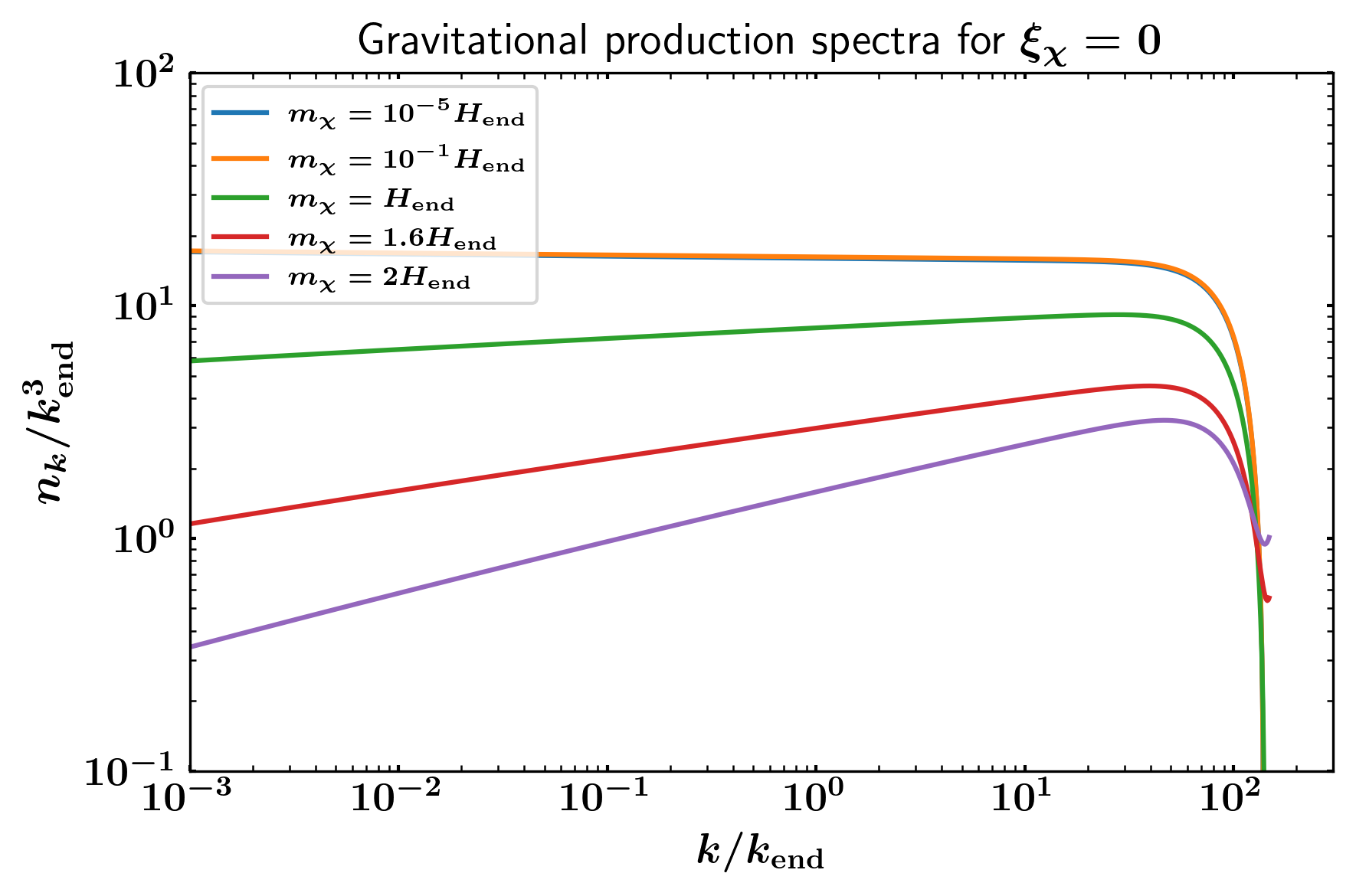}
\caption{The comoving spectrum of gravitational scalar particle production $n_k$ for $\xi_\chi = 0$ as a function of rescaled horizon modes $k/k_\mathrm{end}$. One can see that the spectrum is nearly scale-invariant for masses $m_\chi \lesssim H_\mathrm{end}$ and blue-tilted behavior for larger masses}
   \label{fig8}
\end{figure}

The case with conformal coupling  $\xi_\chi = 1/6$ exhibits a similar behavior to the light mass case $m_\chi \ll H_\mathrm{end}$: in the infrared (IR) region, the spectrum scales as  $n_k \sim k^3$, while in the ultraviolet (UV) region, particle production is exponentially suppressed.

\subsubsection{Minimal coupling $\xi_\chi = 0$}

For the case of $\xi_\chi = 0$, we have a different situation. During inflation, the frequency \eqref{frequency}  is given by

\beq 
    \omega_k^2(t) = \left(\dfrac{k}{a}\right)^2 + e^{-\alpha \phi}m_\chi^2 -\dfrac{9}{4}H^2 + \dfrac{7}{6}\dot{H} - \dfrac{\alpha}{2}V_\phi\,.
    \label{frequency xi = 0}
\enq 

In this case, the derivative potential $V_\phi$ and the Hubble parameter $H^2$ contribute so that the effective mass $m_\mathrm{end}$ is negative. During inflation where $\phi \simeq 5.7 M_p$, $\dot{H} \to 0$ and the mass term $e^{-\alpha \phi} m_\chi^2 \ll H$, so it is negligible. In these cases, the dominant term of \eqref{frequency xi = 0} is $H^2$ if the modes are super-Horizon $k \ll aH$.  Thus, once the modes cross the horizon and for masses $m_\chi \ll H_\mathrm{end}$, this may lead to $\omega_k^2(t) < 0$, resulting in tachyonic instabilities, as mentioned in the discussion following equation \eqref{effective mass} \cite{Markkanen2017}. These instabilities primarily affect the IR modes of the produced particle spectrum, as the modes outside the horizon during inflation experience tachyonic growth. This growth of the modes $X_\mathbf{k}(t)$ outside the horizon is the main driver of the enhanced particle production for minimally coupled scalars compared to conformally coupled scalars. 

\begin{figure}[h]
\centering
\includegraphics[width= 0.89 \textwidth]{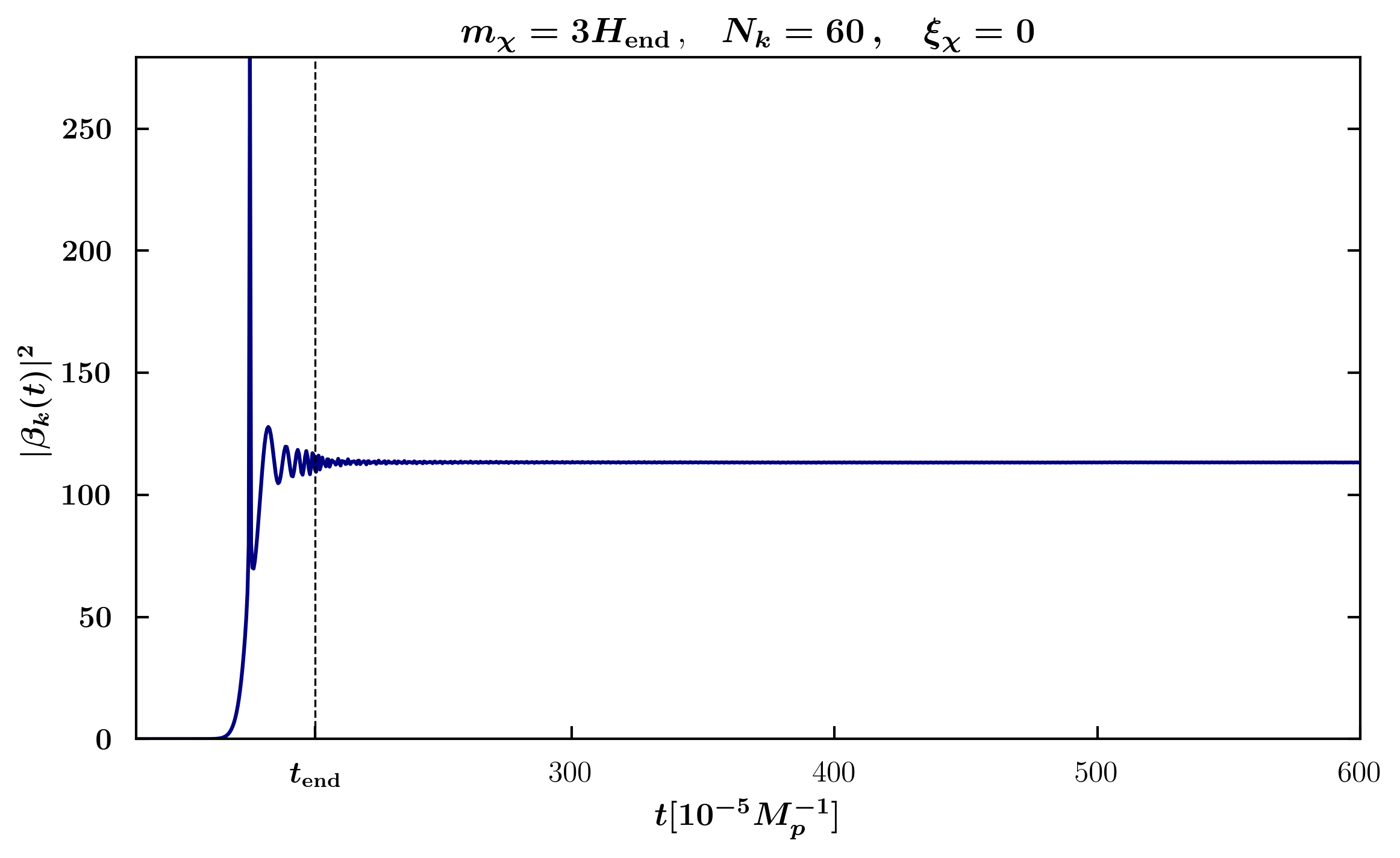}
\caption{The Bogolyubov coefficient $|\beta_k|^2$ as a function of cosmic time $t$ for a spectator mass $m_\chi = 3 H_\mathrm{end}$ and minimal coupling $\xi_\chi =0$.}
   \label{fig9}
\end{figure}

The spectrum of the produced particles is also affected, and its behavior depends on the mass $m_\chi$. To obtain the gravitational spectra, we use the same technique to solve the equation numerically: we choose the $k$-mode inside the horizon, approximately 5 e-folds before it crosses the horizon and show the numerical result for $\beta_k(t)$ at those times when $\omega_k^2 > 0$.   In Fig. \ref{fig8}, we show the spectrum of created particles for minimally coupled masses. We observe that the spectrum is nearly scale-invariant for IR modes and for masses $m_\chi \lesssim H_\mathrm{end}$ before reaching the peak at $k_\mathrm{peak}$. However, as the mass increases, the spectrum tends to be blue-tilted. In the UV region, all modes decay either as a power law or exponentially \cite{Garcia20231}. As the authors mention, the comoving particle number density \eqref{particle number} can have an IR divergence for $m_\chi \ll H_\mathrm{end}$ that is governed by a cutoff scale $k_0$. For $\xi_\chi > 0$, the integral \eqref{particle number} converges and becomes IR insensitive approximately when $\xi_\chi \sim 1/6$.  A clear difference exists between the cases $\xi_\chi = 0$ and $\xi_\chi = 1/6$. In the frequency $\omega_k^2(t)$, the Ricci scalar does not appear for $\xi_\chi = 0$, indicating a significant difference in the behavior of $\omega_k$ for different coupling cases. This is because the Ricci scalar exhibits an oscillatory behavior after inflation due to the oscillations of the fields $\phi$ and $h$. In other words, for the case $\xi_\chi = 1/6$, the Ricci scalar plays an important role in describing the evolution of the modes $X_k(t)$, which translates into a different spectrum of created particles. In the Fig. \ref{fig9} we present the behavior of $|\beta_k(t)|^2$ as a function of time for a mass $m_\chi = 3 H_\mathrm{end}$ and $N_k = 60$ for the minimum coupling case $\xi_\chi =0$. We see the same behavior as in the case $\xi_\chi = 1/6$, a resonant peak near the end of inflation, subsequent oscillations until reaching a stabilized and constant value at $t \gg t_\mathrm{end}$. The difference lies in the order of magnitude of $|\beta_k(t)|^2$, and hence of $n_k$. We can see that the particle production at minimal coupling is much larger than the conformally-coupled case. This is mainly due to the behavior of the spectator field $X_k(t)$ during inflation.

As explained above, the tachyonic instabilities in the case of $\xi_\chi = 0$ cause the modes $X_k(t)$ to grow exponentially. This growth of the super-horizon modes is primarily responsible for the enhanced gravitational particle production for minimally coupled scalars \cite{Garcia20231, kolb2023}.

\section{Relic abundance}

If gravitationally produced particles are sufficiently long-lived, they could survive throughout the entire thermal history of the universe up to the present-day \( t = t_0 \), thereby becoming viable dark matter candidates. These particles might account for only a fraction of the observed dark matter, or they could potentially explain its entirety, as quantified by the parameter \( \Omega_\mathrm{DM} \). For this reason, it is worthwhile to compute both the total number density \( n_\chi \) and the present-day abundance \( \Omega_\chi h^2 \). In this section, we compute the present-day relic abundance of \( \chi \) particles produced gravitationally in the Higgs-\( R^2 \) model, summarize our numerical results, and impose the corresponding relic abundance constraints. The total number of gravitationally produced particles is characterized by two key parameters, depending on the application: either through the comoving particle number density, defined in Eq. \eqref{particle number}, or via the present-day density parameter \( \Omega_\chi \). The present-day particle relic abundance is given by

\beq 
\Omega_\chi h^2 = \dfrac{\rho_\chi(t_0)}{\rho_c}h^2\,.
\enq

\begin{figure}[h]
    \centering
    \includegraphics[width= 0.48 \textwidth]{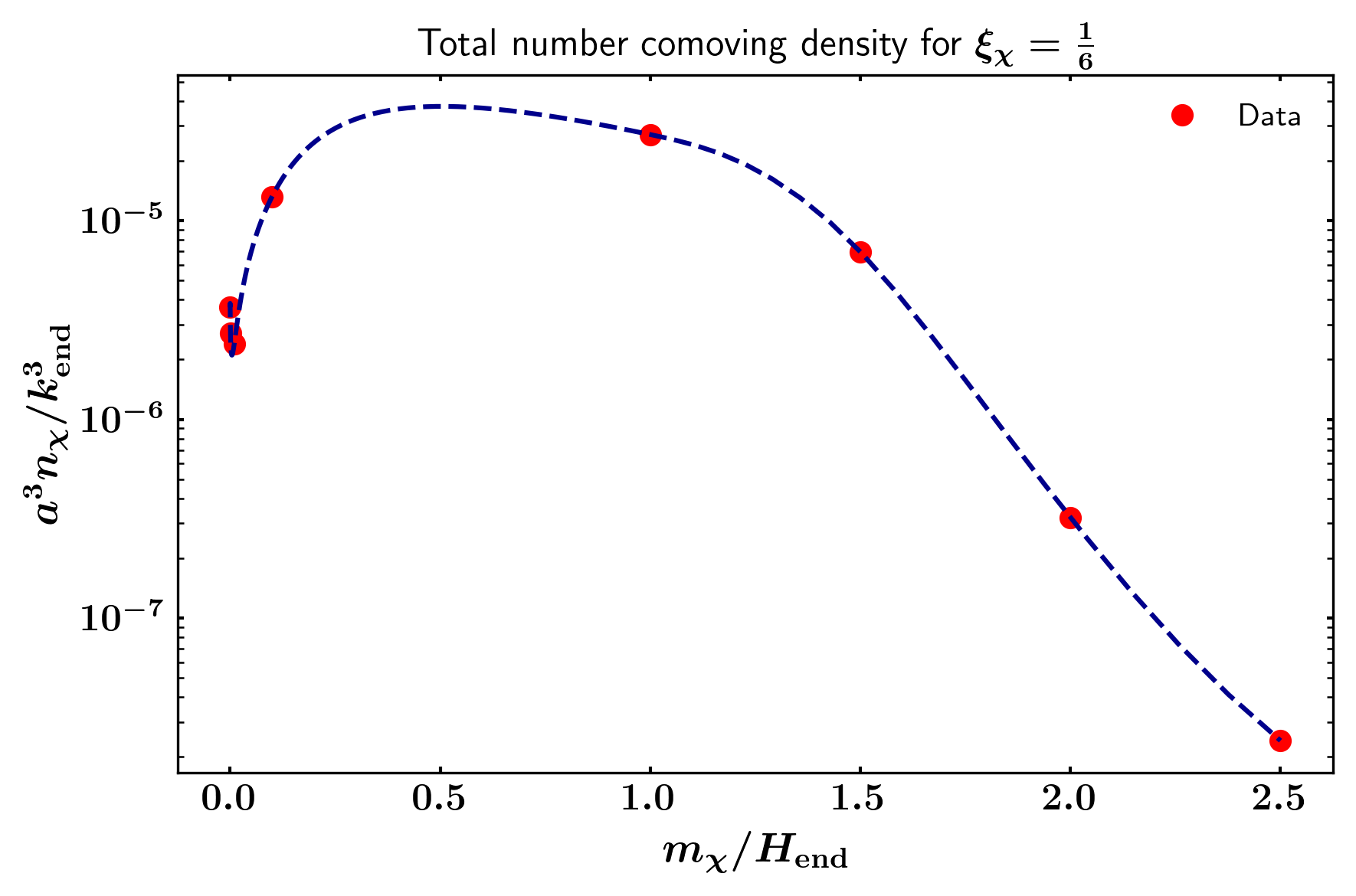}\quad 
    \includegraphics[width= 0.48 \textwidth]{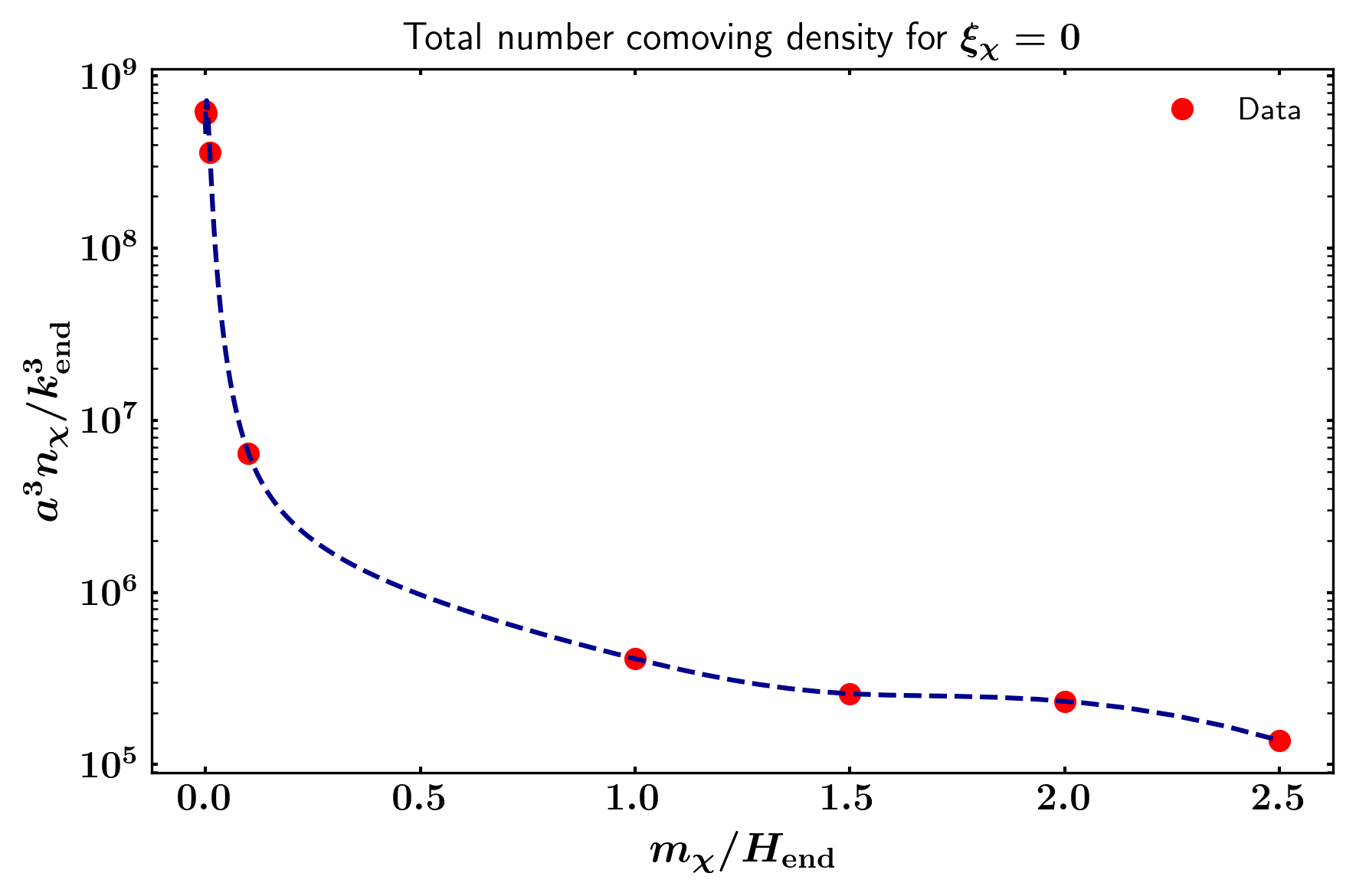}
    \caption{Total comoving number density \( a^3 n_\chi \) in units of \( k_\mathrm{end} \) as a function of the spectator mass \( m_\chi/H_\text{end} \). The left panel shows the result for the conformal coupling case \( \xi_\chi = 1/6 \), while the right panel shows the result for minimal coupling \( \xi_\chi = 0 \).}
    \label{fig10}
\end{figure}

Here, \( \rho_\chi \) is the total energy density of the produced particles, and $\rho_c = 3 M_p^2 H_0^2$  is the critical density of the universe today. Once the asymptotic value of \( n_k(t) \) is known, we can compute \( \Omega_\chi h^2 \) at the present-day. The renormalized total energy density \( \rho_\chi \) is defined as \cite{kolb2023, ema2018production}:

\beq 
\rho_\chi(t) a^4(t) = \int \dfrac{\mathrm d^3 k}{(2\pi)^3}\,\omega_k(t)\,|\beta_k(t)|^2 = \int_0^\infty \dfrac{\dif k}{k}\,\omega_k(t)n_k(t)\,.
\enq 

For non-relativistic modes \( k \ll a\,m_\chi \), which are expected to carry most of the energy at late times, we have:

\beq 
\rho_\chi(t)a^4(t) \simeq \dfrac{m_\chi n_\chi}{\rho_c} \,.
\enq 

That is, the present-day abundance at \( t = t_0 \) can be computed from:

\beq 
\Omega_\chi h^2 \simeq \dfrac{m_\chi n_\chi(t_0)}{\rho_c}h^2 \,.
\label{abundance}
\enq 

It is important to emphasize that the mass term \( m_\chi \) appearing in the frequency \eqref{frequency} is accompanied by an exponential factor \( e^{-\alpha \phi} \), and thus it must be evaluated once \( \phi \) has stabilized after reheating. To perform this calculation, one needs to know the reheating mechanism, the reheating temperature \( T_\mathrm{RH} \), and its duration. However, our particle production calculation is independent of the reheating process, since we assume that the scalar field is a spectator. Reheating should take into account the production of gauge bosons \( W \) and \( Z \) due to Higgs decay and possible resonances.

First, we show the total comoving number density $a^3 n_\chi$ as a function of $m_\chi$ for both $\xi_\chi = 0$ and $\xi_\chi = 1/6$ in Fig. \ref{fig10}. In both cases, we observe a decrease in the number of produced particles as the mass increases, as expected. The superheavy case \( m_\chi \gg H_\text{end} \) is particularly interesting for minimal coupling. In the right panel of Fig. \ref{fig10}, we observe a significant enhancement in the comoving number density \( a^3 n_\chi \) over the explored mass range, reaching a maximum of order \( 10^5 \) for \( m_\chi/H_\text{end} \sim 2.5 \), which corresponds to \( m_\chi \simeq 1.5 \times 10^{13}\,\text{GeV} \). Although this mass is not strictly in the \( m_\chi \gg H_\text{end} \) regime, as the mass increases further, \( a^3 n_\chi \) is expected to decay. However, for a more detailed study of the $\xi_\chi = 0$ case, it is necessary to extend the mass range to the superheavy case, which is computationally more demanding, and requires a much finer numerical analysis. This is left for future work. The conformally coupled case \( \xi_\chi = 1/6 \) follows the expected behavior: it exhibits a peak around \( m_\chi \sim H_\text{end} \), and for heavier spectator masses \( m_\chi \gtrsim H_\text{end} \), gravitational production becomes suppressed.

To calculate the $\Omega_\chi$ abundance, it is necessary to understand the mechanism of the reheating in the model. As mentioned, our work is independent of this process. There are several attempts to explain ths epoch in the Higgs-$R^2$ model. For example, in \cite{BEZRUKOV2019657}, the non-perturbative production of $W $ bosons and Higgs boson self-production after inflation is analyzed. This allows for the computation of relevant observables such as the reheating temperature $T_\mathrm{rh}$. The authors conclude that the dominant post-inflationary dynamics of the model is driven by tachyonic instabilities in the effective masses of the gauge bosons, and to a lesser extent, by the tachyonic mass of the Higgs boson itself. This tachyonic regime leads to a cosmologically instantaneous preheating, which allows the post-inflationary evolution of the model to be fully determined.
In light of these results, we can provide an illustrative analytical estimate for the dark matter abundance $\Omega_\chi h^2$ of the spectator field $\chi$, based on the comoving number density $n_\chi$ and the reheating temperature $T_\text{RH}$. We evaluate the total comoving number density of produced particles at late times, taken as the final time $t_\text{final}$ of our numerical simulations, where the factor $ a^3 n_\chi$ has reached a constant value. Using the asymptotic value of $n_\chi$, we can calculate the relic abundance $\Omega_\chi h^2$ using the amount of expansion from the end of inflation to today. To do this, we must rescale the scale factor from the end of inflation to the present-day, taking into account the reheating epoch. Taking into account that after inflation, the production of gauge bosons begins due to the decay of the Higgs field \cite{BEZRUKOV2019657}, the expression \eqref{abundance} can be rewritten as
\beq 
    \Omega_\chi h^2 \simeq \dfrac{m_\chi n_\chi(t_0)}{\rho_c} = \dfrac{m_\chi }{3M_p^2 H_0^2}\dfrac{a^3 n_\chi}{k_\text{end}^3}\dfrac{k_\text{end}^3}{a_0^3}\,,
\enq 

where $n_\chi(t_0) = a^3 n_\chi/k_\text{end}^3 = H_\text{end}^3[a^3 n_\chi/k_\text{end}^3] (a_\text{end}/a_0)^3 $ is the comoving number density at the present-day. The factor $[a^3 n_\chi/k_\text{end}^3]$ comes from calculating \eqref{particle number} numerically, see Fig. \ref{fig10}. The ratio of the scale factors $(a_\text{end}/a_0)^3$ corresponds to the amount of expansion from the end of inflation to the present-day. Taking into account that after reheating the entropy density $s$ is conserved, the amount of expansion from the end of inflation to the present time can be rewritten as \cite{long2023, Chung2001}

\begin{figure}[h]
    \centering
    \includegraphics[width= 0.89 \textwidth]{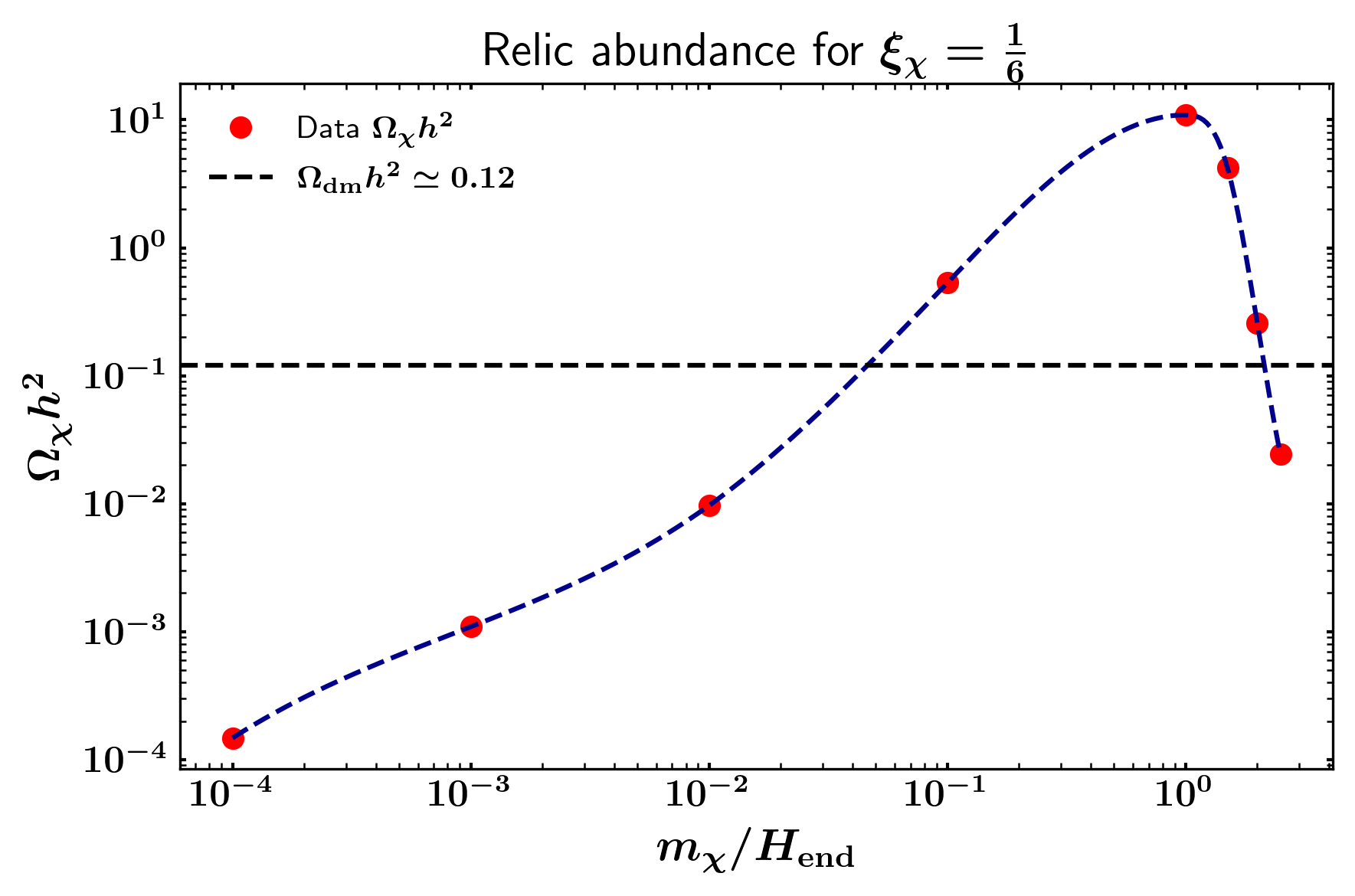}
    \caption{Dimensionless relic abundance \eqref{abundance} of \( \chi \) particles for the case \( \xi_\chi = 1/6 \) as a function of the mass, assuming a reheating temperature \( T_\text{RH} \sim 10^{9}\,\text{GeV} \). The horizontal line indicates the present-day value of the dark matter abundance \( \Omega_\text{CDM} h^2 \sim 0.12 \).}
    \label{fig11}
\end{figure}

\beq 
    \left(\dfrac{a_\text{end}}{a_0}  \right)^3 = \dfrac{\pi^2 g_* T_0^3}{90 M_p^2 H_\text{end}^2}T_\text{RH}\,.
\enq 

Therefore, the abundance \eqref{abundance} of the created particles rescaled to the present-day becomes

\beq
    \Omega_\chi h^2 = \left(\dfrac{\pi^2 g_* T_0^3}{270 M_p^4 H_0^2}\right)\left(\dfrac{m_\chi}{H_\text{end}}\right)\left(\dfrac{a^3 n_\chi}{k_\text{end}^3} \right) H_\text{end}^2\,T_\text{RH}\,,
    \label{abundance rescaled}
\enq 

where $g_* \approx 3.91$, $T_0 = 0.234 \times 10^{-12}\,\text{GeV}$ and $H_0 = 2.133 \times 10^{-42}\,\text{GeV}$ are the numerical values in $\text{GeV}$ of the temperature and the Hubble constant at the present-day. Thus, the numerical factor of $\Omega_\chi h^2$ is 

\beq 
\dfrac{\pi^2 g_* T_0^3}{270 M_p^4 H_0^2} \approx 0.1154 \times 10^{-28}\,\text{GeV}^{-3}
\enq 

The expression \eqref{abundance rescaled} allows us to calculate the relic abundance from the numerical calculation of the comoving number density $a^3 n_\chi$, and as a function of the mass $m_\chi$ and the reheating temperature $T_\text{RH}$.  For example, by fixing $T_\mathrm{RH}$, one can estimate the dependence of $\Omega_\chi$ on the mass $m_\chi$; alternatively, by fixing the ratio $m_\chi/H_\text{end}$, the abundance becomes proportional to $a^3 n_\chi/k_\mathrm{end}^3$, isolating the effect of particle production. Since the reheating temperature is a fixed prediction of this model, we focus on exploring the dependence of $\Omega_\chi$ on the mass $m_\chi$. In Fig. \ref{fig11}, we present the relic abundance of \( \chi \) particles at the present-day as a function of their mass for \( \xi_\chi = 1/6 \), using Eq. \eqref{abundance rescaled} and $T_\text{RH} \sim 10^9$ GeV.
From the figure we find that the maximum allowed mass for the spectator field to account for the observed value \( \Omega_\text{dm} h^2 \) without overproduction is \( m_\chi \simeq 10^{12}\,\text{GeV} \), consistent with other results in the literature for the conformally coupled case \cite{Gorbunov2012, cembranos2020}, although the exact upper bound may vary depending on the specific production mechanism considered. We also observe that for light particles $m_\chi \ll H_\text{end}$, the abundance decreases. This is because in this regime the gravitational production is inefficient due to the conformal coupling, which suppresses the amplification of modes during the expansion of the Universe. Although these masses fail to explain all of the observed dark matter, they may contribute a subdominant fraction.  Therefore, the allowed masses are in the range $m_\chi \lesssim 10^{12}\,\text{GeV}$ for the conformal case, considering only non-perturbative dark matter production and heavy particles $m_\chi \gtrsim H_\text{end}$.

\section{Discussion and remarks}\label{sec12}

In this paper, we have investigated the gravitational and non-perturbative production of free scalar particles, $\chi$, in Higgs-$R^2$ inflation. We started with a free spectator field, $\chi$, interacting only through the gravitational term $\xi_\chi R$ in the Jordan frame. By applying a conformal transformation to transition into the Einstein frame, direct couplings between the $\chi$ field and the scalaron $\phi$ naturally emerged. Within the context of inflationary dynamics in the Einstein frame, we derived the equation of motion for the $\chi$ field, accounting for its non-minimal coupling to the scalar curvature $R$. We numerically solved this system of equations, imposing Bunch-Davies conditions on the $\chi$ field five e-folds before the modes crossed the horizon, to determine the mode functions of $\chi$. From these solutions, we extracted the Bogolyubov coefficient $\beta_k(t)$ to compute the comoving particle density and visualize the resulting spectrum. We explored two distinct scenarios for the non-minimal coupling parameter $\xi_\chi$, considering different mass ranges. For $\xi_\chi = 1/6$, we observed efficient particle production at the end of inflation, driven by the term $e^{-\alpha\phi}m_\chi^2$, which becomes dominant at this stage. The resulting power spectrum exhibited a blue-tilted behavior, indicating that particle production was dominated by modes near the Hubble radius $(aH)^{-1}$ at the end of inflation, while shorter-wavelength modes were suppressed. As the mass increased in the $\xi_\chi = 1/6$ case, particle production tended to decrease due to the increase in the effective mass. In contrast, for $\xi_\chi = 0$, the spectrum remained nearly scale-invariant for masses $m_\chi \lesssim H_\mathrm{end}$, with a blue tilt emerging for larger masses $m_\chi \gtrsim H_\mathrm{end}$. For light masses and $\xi_\chi = 0$, we found that the comoving number of particles created was significantly higher compared to the case of conformally coupled particles, primarily due to the tachyonic growth of the modes $X_k(t)$.

While we did not conduct a comprehensive scan of the parameter space for Higgs-$R^2$ inflation or the $\chi$ field, we focused on a benchmark scenario consistent with current CMB observations. We found that non-perturbative production of spectator scalars with conformal coupling can yield viable dark matter candidates for masses up to $m_\chi \sim 10^{12}\,\mathrm{GeV}$. Lighter masses fail to account for the full relic abundance but may still contribute a subdominant component. The superheavy regime $m_\chi \gg H_\text{end}$ is particularly intriguing, involving a different production mechanism driven by adiabaticity violation, but is left for future work due to its computational demands. 

Our results underscore the importance of gravitational production mechanisms in extended inflationary scenarios and highlight the viability of free spectator scalars as dark matter candidates, even in the absence of direct couplings to the Standard Model.

There remain several directions for future research, including the production of spin-$1/2$ and spin-$1$ dark matter, the production of gauge bosons $W^\pm_\mu$, $Z_\mu$, and the implications of particle production on primordial non-Gaussianities in the CMB. It would also be interesting to study the isocurvature dark matter constraints for light scalar field masses in this model. We leave these intriguing topics for future investigation.

\bmhead{Acknowledgements}

We are very grateful to Dr. Marcos García for his insightful comments and guidance, which were instrumental in the development of the numerical code used in this work. We also thank Dr. Andrew Long for generously providing an example of numerical code that served as a valuable reference during the initial stages of this project. We would like to express our sincere appreciation to the anonymous referee for their constructive suggestions and careful reading of the manuscript, which helped us significantly improve the clarity and quality of this work.

The work of F.P. has been supported by CONACyT-México through the doctoral scholarship 803062. This research was also partially supported by CONACyT-México through the grant CBF2023-2024-193.

\bibliography{main}

\end{document}